\documentclass[9pt]{article}
\usepackage[margin=1in]{geometry}

\usepackage[english]{babel}
\usepackage[utf8]{inputenc}
\usepackage{url}  
\usepackage{graphicx}
\usepackage{caption}
\usepackage{subcaption}
\usepackage{soul}
\usepackage{float}
\usepackage{amsmath}

\usepackage[table]{xcolor}

\usepackage{tikz}
\usetikzlibrary{shapes}

\usepackage{authblk}
\usepackage[nottoc]{tocbibind}

\usepackage[normalem]{ulem}
\usepackage{colortbl}
\usepackage{booktabs}
\usepackage{longtable}

\newcolumntype{L}[1]{>{\raggedright\arraybackslash}p{#1}}
\usepackage{adjustbox}
\usepackage{makecell}

\newcolumntype{C}[1]{>{\centering\arraybackslash}p{#1}}  
\usepackage{array}

\definecolor{lightgrey}{RGB}{239,239,239}
\definecolor{mediumgrey}{RGB}{111,111,111}
\definecolor{headerblue}{RGB}{180,210,255}
\definecolor{mediumblue}{RGB}{61,133,198}
\definecolor{lightgreen}{RGB}{226,244,235}
\definecolor{lightpink}{RGB}{223,168,189}
\definecolor{lightblue}{RGB}{98,117,162}
\definecolor{blueteal}{RGB}{164,209,215}
\definecolor{lightviolet}{RGB}{167,158,186}
\definecolor{white}{RGB}{255,255,255}
\definecolor{green0}{RGB}{255,255,255}
\definecolor{green1}{RGB}{235,255,235}
\definecolor{green2}{RGB}{215,245,215}
\definecolor{green3}{RGB}{190,235,190}
\definecolor{green4}{RGB}{160,220,160}
\definecolor{green5}{RGB}{130,210,130}
\definecolor{green6}{RGB}{100,195,100}
\definecolor{green7}{RGB}{70,180,70}
\definecolor{green8}{RGB}{40,160,40}
\definecolor{green9}{RGB}{20,140,20}
\definecolor{greenA0}{RGB}{60,160,60}
\definecolor{greenA1}{RGB}{120,200,120}
\definecolor{greenA2}{RGB}{180,235,180}
\definecolor{greenA3}{RGB}{220,255,220}
\definecolor{greenA4}{RGB}{255,255,255}
\definecolor{pink0}{RGB}{255,102,153}
\definecolor{pink1}{RGB}{255,140,180}
\definecolor{pink2}{RGB}{255,185,210}
\definecolor{pink3}{RGB}{255,220,230}
\definecolor{pink4}{RGB}{255,255,255}

\newcolumntype{L}[1]{>{\raggedright\arraybackslash}p{#1}} % left-aligned
\newcolumntype{C}[1]{>{\centering\arraybackslash}p{#1}}   % centered
\usepackage{longtable}
\usepackage[table]{xcolor}
\usepackage{array}
\newcolumntype{C}[1]{>{\centering\arraybackslash}p{#1}}
\definecolor{darkpink}{RGB}{179, 50, 107}

\renewcommand{\arraystretch}{1.8}

\newcommand{\percentColorCell}[1]{%
  \begingroup
  \def\val{#1}%
  \ifdim \val pt < 50pt
    \ifdim \val pt < 10pt 
      \cellcolor{greenA0}#1\% 
    \else
      \ifdim \val pt < 25pt 
        \cellcolor{greenA1}#1\% 
      \else
        \ifdim \val pt < 40pt 
          \cellcolor{greenA2}#1\% 
        \else
          \cellcolor{greenA3}#1\% 
        \fi
      \fi
    \fi
  \else
    \ifdim \val pt = 50pt 
      \cellcolor{greenA4}#1\% 
    \else
      \ifdim \val pt < 60pt 
        \cellcolor{pink3}#1\% 
      \else
        \ifdim \val pt < 70pt 
          \cellcolor{pink2}#1\% 
        \else
          \ifdim \val pt < 80pt 
            \cellcolor{pink1}#1\% 
          \else
            \cellcolor{pink0}#1\% 
          \fi
        \fi
      \fi
    \fi
  \fi
  \endgroup
}

\newcommand{\colorPercentCell}[1]{%
  \begingroup
  \def\val{#1}%
  \ifdim \val pt < 50pt
    \ifdim \val pt < 10pt 
      \cellcolor{greenA0}#1\% 
    \else
      \ifdim \val pt < 25pt 
        \cellcolor{greenA1}#1\% 
      \else
        \ifdim \val pt < 40pt 
          \cellcolor{greenA2}#1\% 
        \else
          \cellcolor{greenA3}#1\% 
        \fi
      \fi
    \fi
  \else
    \ifdim \val pt = 50pt 
      \cellcolor{greenA4}#1\% 
    \else
      \ifdim \val pt < 60pt 
        \cellcolor{pink3}#1\% 
      \else
        \ifdim \val pt < 70pt 
          \cellcolor{pink2}#1\% 
        \else
          \ifdim \val pt < 80pt 
            \cellcolor{pink1}#1\% 
          \else
            \cellcolor{pink0}#1\% 
          \fi
        \fi
      \fi
    \fi
  \fi
  \endgroup
}

\title{Mapping the interaction between science and misinformation in COVID-19 tweets}

\author[1]{Lucila G. Alvarez-Zuzek
\thanks{lalvarezzuzek@fbk.eu}}
\author[2]{Juan P. Bascur}
\author[1,3]{Anna Bertani}
\author[1]{Riccardo Gallotti}
\author[2]{Vincent A. Traag}
\affil[1]{Complex Human Behaviour Lab, Fondazione Bruno Kessler, Trento, Italy}
\affil[2]{Centre for Science and Technology Studies (CWTS), Leiden University, Leiden, The Netherlands}
\affil[3]{Department of Engineering and Computer Science, University of Trento, Trento, Italy}

\begin{document}

\maketitle

\begin{abstract}
    
During the COVID-19 pandemic, scientific knowledge evolved rapidly, accompanied by a surge of misinformation, labelled an infodemic by the WHO. In this context, we study the interaction between science and misinformation on Twitter (now X) using a database of $\sim407M$ COVID-19-related tweets. We classify URL reliability with Media Bias/Fact Check and used Altmetric data to identify scientific publications. We find that among $\sim1.2M$ users who shared science, $45\%$ also shared unreliable content. Scientific papers circulated by these users were more often preprints, slightly more likely to be retracted, less cited, and published in lower-impact journals. Our findings indicate misinformation is not driven by a lack of exposure to science but instead raise critical questions about open science practices, particularly the role of preprints in amplifying misleading narratives. Our results underscore the importance of proactive scientific engagement on social media in countering misinformation and reinforcing trust in science during global crises.
    
\end{abstract}

%%%%%%%%%%%%%%%%%%%%%%%%%%%%%%%%%%%%%%%%%%%%%%%%%%%%%%%%%%%%%%%%%%%%
%%%%%%%%%%%%%%%%%%%%%%%%%%%%%%%%%%%%%%%%%%%%%%%%%%%%%%%%%%%%%%%%%%%%
%%%%%%%%%%%%%%%%%%%%%%%%%%%%%%%%%%%%%%%%%%%%%%%%%%%%%%%%%%%%%%%%%%%%
%%%%%%%%%%%%%%%%%%%%%%%%%%%%%%%%%%%%%%%%%%%%%%%%%%%%%%%%%%%%%%%%%%%%
%%%%%%%%%%%%%%%%%%%%%%%%%%%%%%%%%%%%%%%%%%%%%%%%%%%%%%%%%%%%%%%%%%%%
%%%%%%%%%%%%%%%%%%%%%%%%%%%%%%%%%%%%%%%%%%%%%%%%%%%%%%%%%%%%%%%%%%%%
%%%%%%%%%%%%%%%%%%%%%%%%%%%%%%%%%%%%%%%%%%%%%%%%%%%%%%%%%%%%%%%%%%%%
%%%%%%%%%%%%%%%%%%%%%%%%%%%%%%%%%%%%%%%%%%%%%%%%%%%%%%%%%%%%%%%%%%%%
%%%%%%%%%%%%%%%%%%%%%%%%%%%%%%%%%%%%%%%%%%%%%%%%%%%%%%%%%%%%%%%%%%%%
%%%%%%%%%%%%%%%%%%%%%%%%%%%%%%%%%%%%%%%%%%%%%%%%%%%%%%%%%%%%%%%%%%%%
%%%%%%%%%%%%%%%%%%%%%%%%%%%%%%%%%%%%%%%%%%%%%%%%%%%%%%%%%%%%%%%%%%%%
%%%%%%%%%%%%%%%%%%%%%%%%%%%%%%%%%%%%%%%%%%%%%%%%%%%%%%%%%%%%%%%%%%%%
%%%%%%%%%%%%%%%%%%%%%%%%%%%%%%%%%%%%%%%%%%%%%%%%%%%%%%%%%%%%%%%%%%%%
%%%%%%%%%%%%%%%%%%%%%%%%%%%%%%%%%%%%%%%%%%%%%%%%%%%%%%%%%%%%%%%%%%%%
%%%%%%%%%%%%%%%%%%%%%%%%%%%%%%%%%%%%%%%%%%%%%%%%%%%%%%%%%%%%%%%%%%%%
\section*{Introduction}

Misinformation can be seen as the antithesis of science. Whereas science tries to uncover truths, misinformation covers up truths. At the same time, distinguishing false narratives from fact is not always straightforward~\cite{boumans_fostering_2025}, particularly as scientific insights evolve, as became profusely clear during the COVID-19 pandemic. Defining misinformation in the context of science is challenging, as also raised in a recent report by the US Academy of Sciences~\cite{nas_understanding_2024}, who settle on the following definition:
\begin{quote}
    Misinformation about science is information that asserts or implies claims that are inconsistent with the weight of accepted scientific evidence at the time (reflecting both quality and quantity of evidence).
    Which claims are determined to be misinformation about science can evolve over time as new evidence accumulates and scientific knowledge regarding those claims advances.
    \cite[p. 32]{nas_understanding_2024}
\end{quote}
Misinformation plays a particularly prominent role in the engagement of the broader public with science, and this calls for attention to how science interacts with misinformation. 

The interaction of the broader public with science is traditionally understood in science communication as a linear process in which scientists produce knowledge, science journalists disseminate knowledge in accessible formats, and the public receives information. This linear science communication model shows clear shortcomings, and studies have shown clear limitations of the ``deficit model'' that underpins the idea that communicating science improves public scientific knowledge~\cite{scheufele_communicating_2013}. Instead, social actors in science and technology play a more complex role in shaping the interaction between science and society~\cite{irwin_misunderstanding_1996}. The emergence of social media has further changed the nature of science communication. There has been a shift in who communicates science~\cite{hansen_changing_2016}, opening it up to a broader range of people. In addition, audiences can select information from a wide array of sources, including science, as it is increasingly openly available. This is not just about accessing science, but also about accessing unsettled truth-claims \cite{van_schalkwyk_amplification_2019}, rendering visible some of the uncertainties and controversies that are part of science~\cite{latour_science_1988,venturini_controversy_2021}.
Finally, a decline in science journalism has led to less critical use of press releases~\cite{schafer_how_2017}.

During the COVID-19 pandemic, many of these developments became more apparent.
Misinformation was very visible during the pandemic~\cite{gallotti2020assessing} to the extent that WHO spoke of an ``infodemic''. For instance, exposure to misinformation seems to be associated with lower vaccine intentions~\cite{singh_misinformation_2022,greene_quantifying_2021}, although the implications of misinformation are not fully understood~\cite{adams_why_2023}. Indeed, some suggest that the ``infodemic'' was not really a novel phenomenon \cite{scheufele_misinformed_2021} and requires a more nuanced approach~\cite{krause_infodemic_2022}. In contrast with the ``deficit model'', there are clear instances of use of science in anti-vaccine websites~\cite{moran_what_2016}, and social media~\cite{van_schalkwyk_amplification_2019}. More generally, what topics become public controversies may not be related to misinformation~\cite{kahan_sources_2017}.

Beyond misinformation about science, there are also potential problems within science itself~\cite{west_misinformation_2021}, including problems of predatory journals, and questionable research practices~\cite{swire-thompson_reducing_2022}. One notable issue over the past decade is the attention on problems of replicability~\cite{nosek_replicability_2022} with ensuing potential implications for trust in science~\cite{hendriks_replication_2020}. In addressing some of these issues, open science practices are considered to play an important role~\cite{cole_societal_2024,klebel_academic_2025}. Indeed, open science may increase trust in science~\cite{rosman_open_2022,song_trusting_2022}. However, it is not clear whether trust is necessarily problematic, as it generally remains high, although it varies across demographics and countries~\cite{cologna2025trust}, with some political cleavages, especially in the US~\cite{milkoreit_rapidly_2024,gauchat_politicization_2012}.

However, some open science practices may have potentially detrimental effects in the interaction with misinformation. In particular, some scholars raise concerns over potential problems with preprints~\cite{sheldon_preprints_2018}, some going as far as to talk about ``preprint wars''~\cite{silva_preprint_2017}. During the COVID-19 pandemic, there were calls for greater attention on preprints in the context of misinformation, and to consider alternative approaches~\cite{heimstadt_between_2020,ravinetto_preprints_2021}. Some of these potential problems may have been compounded by the faster publication cycle during the COVID-19 pandemic~\cite{horbach_pandemic_2020}. Preprints were particularly reported on in the media during the COVID-19 pandemic~\cite{fleerackers_unreviewed_2024,alperin_stark_2024}. Some problematic preprints received much media attention~\cite{west_misinformation_2021}, even though some high-profile articles published in reputable journals were also problematic and retracted~\cite{ledford_high-profile_2020}.

We here investigate the interaction between misinformation and science on Twitter (now X) in the context of the COVID-19 pandemic. In particular, we study over 400 million COVID-19 related tweets collected during the pandemic from 2020 to 2023 by the COVID-19 Infodemics Observatory~\cite{gallotti2020assessing}. We consider for each tweet whether it refers to a URL, and classify the reliability of the URL on the basis of MediaBias/FactCheck~\cite{MediaBiasFactCheck}. We consider as reliable sources any URL domains that are classified as mainstream media or science, and consider as unreliable sources URL domains that are classified as satire, clickbait, political news, fake/hoax and conspiracy/junk science. Unreliable sources are more likely to contain misinformation\footnote{Misinformation can also include disinformation, with an explicit intent to deceive. We cannot infer the intent on the basis of our data, and hence cannot distinguish between misinformation and disinformation here.}. Some URL domains do not allow for a clear classification of reliability, and we label them as unknown reliability. In addition to the source-based classification on the basis of MediaBias/FactCheck, we also rely on data from Altmetric\footnote{\url{https://www.altmetric.com/}}, to determine whether a tweet refers to a scientific publication. We also classify those tweets as referring to science, and hence also as a reliable source. If a publication is mentioned in a Tweet, we use data from OpenAlex\footnote{\url{https://openalex.org/}} to collect additional bibliometric details, including whether it is open access, a preprint, retraction status and several indicators related to citation impact. In addition, we identify which Twitter users are scientists on the basis of previous work~\cite{costas2020large,mongeon2023open}.
See Methods Section for details. We provide an interactive dashboard for exploring some of our results in more detail\footnote{Available from \url{https://public.tableau.com/views/UnMiSSedDashboard/Landingpage}}.

We investigate the interaction between misinformation and science, specifically how science is used in the context of misinformation. To this end, we analyse the extent to which users share both scientific content and unreliable sources, which allows us to identify publications that are more shared in unreliable contexts and examine their bibliometric characteristics. Finally, we also study how scientists engage in online discussions during COVID-19 and how their participation relates to the spread of misinformation.

%%%%%%%%%%%%%%%%%%%%%%%%%%%%%%%%%%%%%%%%%%%%%%%%%%%%%%%%%%%%%%%%%%%%
%%%%%%%%%%%%%%%%%%%%%%%%%%%%%%%%%%%%%%%%%%%%%%%%%%%%%%%%%%%%%%%%%%%%
%%%%%%%%%%%%%%%%%%%%%%%%%%%%%%%%%%%%%%%%%%%%%%%%%%%%%%%%%%%%%%%%%%%%
%%%%%%%%%%%%%%%%%%%%%%%%%%%%%%%%%%%%%%%%%%%%%%%%%%%%%%%%%%%%%%%%%%%%
%%%%%%%%%%%%%%%%%%%%%%%%%%%%%%%%%%%%%%%%%%%%%%%%%%%%%%%%%%%%%%%%%%%%
%%%%%%%%%%%%%%%%%%%%%%%%%%%%%%%%%%%%%%%%%%%%%%%%%%%%%%%%%%%%%%%%%%%%
%%%%%%%%%%%%%%%%%%%%%%%%%%%%%%%%%%%%%%%%%%%%%%%%%%%%%%%%%%%%%%%%%%%%
%%%%%%%%%%%%%%%%%%%%%%%%%%%%%%%%%%%%%%%%%%%%%%%%%%%%%%%%%%%%%%%%%%%%
%%%%%%%%%%%%%%%%%%%%%%%%%%%%%%%%%%%%%%%%%%%%%%%%%%%%%%%%%%%%%%%%%%%%
%%%%%%%%%%%%%%%%%%%%%%%%%%%%%%%%%%%%%%%%%%%%%%%%%%%%%%%%%%%%%%%%%%%%
%%%%%%%%%%%%%%%%%%%%%%%%%%%%%%%%%%%%%%%%%%%%%%%%%%%%%%%%%%%%%%%%%%%%
%%%%%%%%%%%%%%%%%%%%%%%%%%%%%%%%%%%%%%%%%%%%%%%%%%%%%%%%%%%%%%%%%%%%
%%%%%%%%%%%%%%%%%%%%%%%%%%%%%%%%%%%%%%%%%%%%%%%%%%%%%%%%%%%%%%%%%%%%
%%%%%%%%%%%%%%%%%%%%%%%%%%%%%%%%%%%%%%%%%%%%%%%%%%%%%%%%%%%%%%%%%%%%
%%%%%%%%%%%%%%%%%%%%%%%%%%%%%%%%%%%%%%%%%%%%%%%%%%%%%%%%%%%%%%%%%%%%
\section*{Results}

%%%%%%%%%%%%%%%%%%%%%%%%%%%%%%%%%%%%%%%%%%%%%%%%%%%%%%%%%%%%%%%%%%%%
%%%%%%%%%%%%%%%%%%%%%%%%%%%%%%%%%%%%%%%%%%%%%%%%%%%%%%%%%%%%%%%%%%%%
%%%%%%%%%%%%%%%%%%%%%%%%%%%%%%%%%%%%%%%%%%%%%%%%%%%%%%%%%%%%%%%%%%%%
%%%%%%%%%%%%%%%%%%%%%%%%%%%%%%%%%%%%%%%%%%%%%%%%%%%%%%%%%%%%%%%%%%%%
%%%%%%%%%%%%%%%%%%%%%%%%%%%%%%%%%%%%%%%%%%%%%%%%%%%%%%%%%%%%%%%%%%%%
\subsection*{Overall statistics}

Over time, the total number of tweets collected every day in association with COVID-19 (see methods) decreases, mirroring the progression of the pandemic. Although unreliable tweets follow a similar pattern, the number of tweets referencing scientific research has steadily increased. This suggests that science became progressively more relevant in COVID-19-related debates, reaching comparable levels by the second half of 2022 (see Supplementary Fig.~\ref{fig:timeline}). To better understand the potential diffusion of tweets, we consider the number of followers of users posting a message as an indicator of the exposure of their tweets \cite{gallotti2020assessing} (see Methods Section for a more formal definition). When considering the exposure for both types, science remains lower in comparison with unreliable sources throughout the period.

Our data set indicates that tweets that reference reliable sources are more prevalent ($14.83\%$) compared to tweets referencing unreliable sources ($5.67\%$). An even smaller percentage of tweets refer to science ($1.91\%$). Most users share tweets without any domain, thus without a reliability score ($57.91\%$), or containing a domain for which the reliability is unknown ($21.59\%$). 

Of all users, $28.36\%$ shared at least one tweet containing reliable sources, while $13.81\%$ shared at least one tweet with unreliable sources, meaning twice as many users shared a reliable source as an unreliable source. We refer to users sharing at least one tweet containing unreliable sources as unreliable users, and as reliable users otherwise. When accounting for the number of followers, we observe that reliable users have, on average, more followers than unreliable users ($43,231$ vs. $34,243$). Scientific content is shared by users with comparatively fewer followers, averaging $6,837$.

The relative exposure is considerably higher for tweets with reliable sources (constituting $18.53\%$ of the total exposure of all tweets) compared to those from unreliable sources ($5.61\%$), with tweets on science having a very low exposure ($0.38\%$).
\begin{table}[H]
    \centering
    \rowcolors{2}{gray!10}{white}
    \begin{tabular}{l r r r r}
    \rowcolor{blueteal!50}
                  & Science & Reliable & Unreliable & Untrustworthy \\
    Science       & -       & -        & 46.09 & 22.33 \\
    Reliable      & -       & -        & 30.65 & 10.02 \\
    Unreliable    & 19.64   & 62.92    & -     & -     \\
    Untrustworthy & 35.21   & 76.08    & -     & -     \\
    \end{tabular}
    \caption{{\bf Overlap between users sharing content with indicated categories.} Each row shows the percentage of users who shared content (i.e. at least one tweet) in that row who also share content (i.e. at least one tweet) in the category in the column. Note that percentages do not add up to 100\% because users can share multiple types of content.}
    \label{table:overlap-categories}
\end{table}
We consider whether users share multiple types of content (Table~\ref{table:overlap-categories}). We find that $46.09\%$ of the users who share science also share unreliable sources. %, which is substantially more than users who shared reliable sources in general ($30.65\%$).
Conversely, $19.64\%$ of the users who share unreliable sources also share science. In addition, the separate categories in Table~\ref{table:overlap-categories} make clear that science is more likely to be engaged by users who also share unreliable sources, with a particular concentration in the most unreliable sources (Untrustworthy). Furthermore, at the user level, we observe a weak correlation between the fraction of tweets containing science and unreliable content, with a correlation coefficient of $R = 0.21$, as shown in Fig.~\ref{fig:correlation_per_user}. In sum, users who share unreliable sources are likely to also share science.

Unpacking the sharing of unreliable sources into their constituent categories, we find that users who shared conspiracy theories ($48.48\%$), satire ($45.28\%$), and fake news ($36.61\%$) were the most likely to also share science content.

Sources classified as fake news and conspiracy theories are considered to be most harmful, and we here refer to these categories as ``untrustworthy'' (see the Methods Section for further details). We see that the sharing of science by users who shared untrustworthy sources is relatively high with $35.21\%$ (Table~\ref{table:overlap-categories}). Vice versa, we see that $22.33\%$ of the users who share science also shared untrustworthy sources.

Finally, science is substantially less often shared by users also sharing mild-reliability score categories, $e.i.$ political media ($22.14\%$) and mainstream media ($15.03\%$), as we can see in Table~\ref{table:pairwise-overlap}.

%%%%%%%%%%%%%%%%%%%%%%%%%%%%%%%%%%%%%%%%%%%%%%%%%%%%%%%%%%%%%%%%%%%%
%%%%%%%%%%%%%%%%%%%%%%%%%%%%%%%%%%%%%%%%%%%%%%%%%%%%%%%%%%%%%%%%%%%%
%%%%%%%%%%%%%%%%%%%%%%%%%%%%%%%%%%%%%%%%%%%%%%%%%%%%%%%%%%%%%%%%%%%%
%%%%%%%%%%%%%%%%%%%%%%%%%%%%%%%%%%%%%%%%%%%%%%%%%%%%%%%%%%%%%%%%%%%%
%%%%%%%%%%%%%%%%%%%%%%%%%%%%%%%%%%%%%%%%%%%%%%%%%%%%%%%%%%%%%%%%%%%%
\subsection*{Characterising publications}

To deepen our understanding of the interaction between science and unreliable sources, we now explore the scientific publications mentioned in tweets in more detail. Specifically, we select a subset of publications shared by at least $100$ unique users, resulting in a total of $7,905$ publications. For each publication, we calculate the proportion of users who also disseminated unreliable content, capturing the extent to which each publication is shared by unreliable users. The median value of this distribution is $76.11\%$, which we used as a threshold to partition this data set into two equally sized groups: i) publications with a percentage of unreliable users below the median, therefore articles shared predominantly by {\it reliable users}, which we term the reliably used publications, and ii) publications with a percentage above the median, thus articles shared predominantly by {\it unreliable users}, which we term the unreliably used publications.

\begin{table}[ht]
    \centering
    \footnotesize
    \renewcommand{\arraystretch}{2} 
    \setlength{\tabcolsep}{8pt}     
    \rowcolors{2}{gray!10}{white}
    \begin{tabular}{>{\raggedright\arraybackslash}p{6cm} >{\centering\arraybackslash}p{3cm} >{\centering\arraybackslash}p{3cm}}
    \rowcolor{blueteal!50}
    \textbf{Metric} & \textbf{Reliably used} & \textbf{Unreliably used} \\
    Average number of citations & 370.12 & 107.69 \\
    Average normalised citations & 11.67 & 5.50 \\    
    Average citations of publication sources & 90.31 & 61.65 \\
    Average normalised citations publication sources & 4.73 & 3.29 \\    
    \% retracted & 0.10\% & 0.20\% \\
    \% open-access & 91.94\% & 93.65\% \\
    \% published in an open-access source & 24.09\% & 45.48\% \\
    \% published in a preprint repository & 10.03\% & 17.21\% \\
    Average number of users per DOI & 447.67 & 555.31 \\
    Average number of Tweets per DOI and user & 1.12 & 1.17 \\
    \end{tabular}
    \caption{{\bf Statistics of the publications}. By selecting a subset of publications shared on Twitter by at least $100$ users, we identified a total of $7,905$ publications and divided them into two equally sized groups that have been reliably, respectively, unreliably used publications.}
    \label{table:statistics_dois}
\end{table}

As shown in Table~\ref{table:statistics_dois}, we find that unreliably used publications tend to be published more often in open-access sources ($93.65\%$ vs. $91.94\%$) and preprint repositories ($17.21\%$ vs $10.03\%$). More generally, the journal impact of unreliably used publications tends to be lower (journal impact $61.65$ vs $90.31$, normalised journal impact $3.29$ vs $4.73$). Similarly, the citation impact of unreliably used publications tends to be lower (mean citations $107.69$ vs $370.12$, mean normalised citations $5.50$ vs $11.67$).

We provide an overview of relevant scientific topics, established on the basis of a fine-grained publication-level classification\footnote{Based on the approach by~\cite{waltman_new_2012}, implemented using the Leiden algorithm~\cite{traag_louvain_2019}, with recent improvements in labelling~\cite{van_eck_open_2024}.} in Table~\ref{fig:topics-dois}. As expected, most of the research topics relate to various aspects of the COVID-19 pandemic, including neurological aspects and modelling the dynamics of the pandemic. Some topics exhibit relatively few unreliably used publications, such as those on mental health, mechanical ventilation, and gender bias in science. Certain topics show a relatively high concentration of unreliably used publications, including, for instance, those on vitamin D, parasitic diseases (primarily related to ivermectin), thrombosis (mainly as potential outcomes of COVID-19 vaccines), and the healthcare workforce (related to potential medical malpractice and medical corruption).

The publications are distributed across a wide range of sources, with a notable concentration in \textit{medRxiv}, \textit{Nature}, and \textit{BMJ} (see Table~\ref{table:sources-dois}). A higher percentage of unreliable used publications appear in journals related to infectious diseases, whereas this tendency is lower in journals focused on medicine. For instance, almost all the publications from the journal {\it Cureus} are used in unreliable contexts, while almost no publications from {\it Intensive Care Medicine} were unreliably used. A notable number of publications unreliably used originate from pre-print repositories, particularly {\it medRxiv} and {\it bioRxiv}, according to the overall higher percentage of pre-prints that are unreliably used. 

\begin{figure}[H]
    \centering
    \includegraphics[width=1\linewidth]{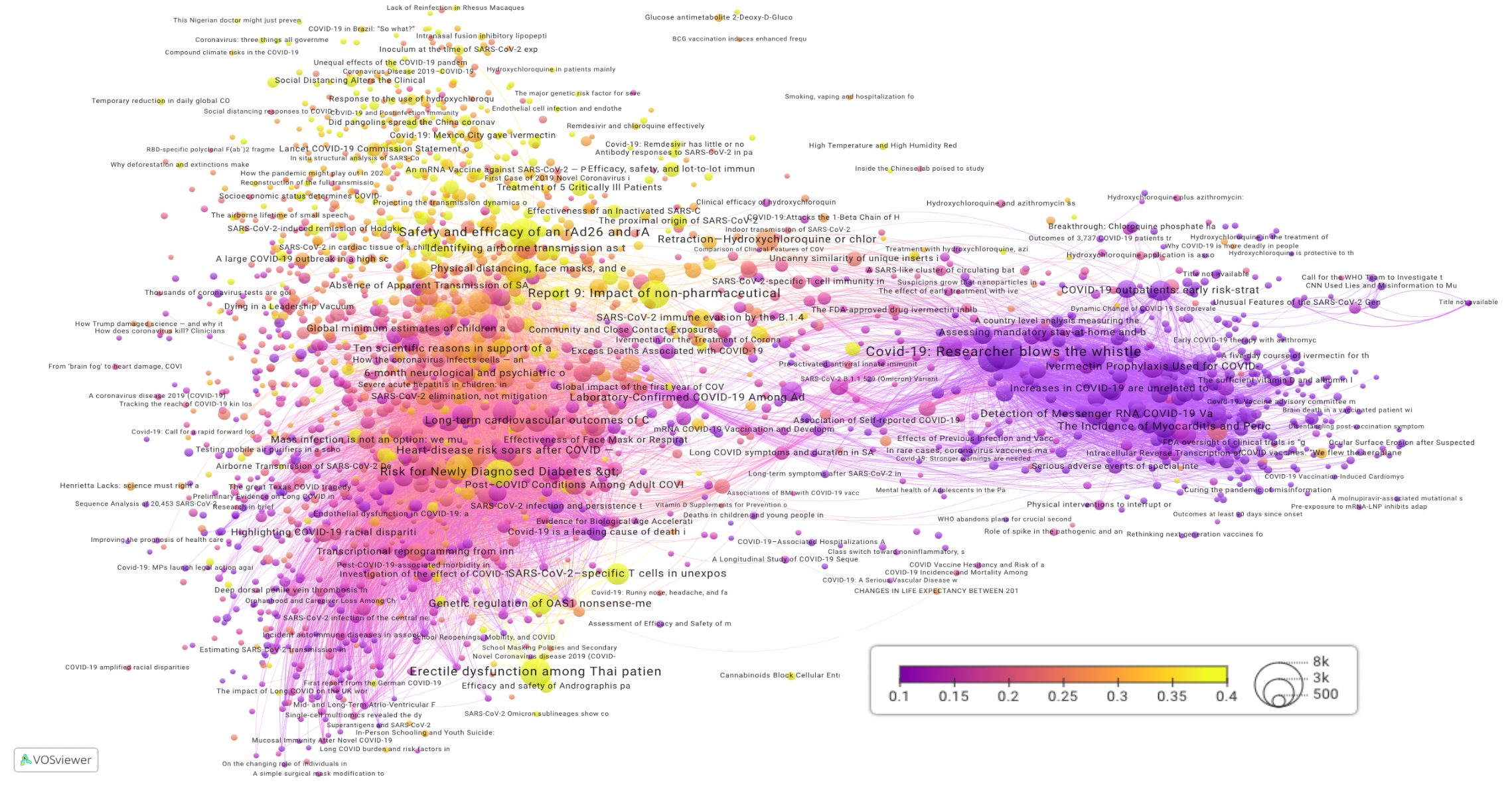}
    \caption{{\bf Network of publications co-shared by reliable and unreliable users}. To examine how scientific publications are discussed on Twitter in more detail, we construct the network of publications co-tweeted by the same user. Specifically, each node represents a scientific article, and edges connect articles that have been tweeted by the same user. Each node represents a scientific publication, and an edge is established between two nodes if both articles were tweeted by at least one user. For the resulting network, we are considering the top $N=2,000$ nodes and $E=990,930$ edges. For clarity of the visualisation, only $8,000$ edges are displayed. Node size represents the number of users who mentioned the corresponding article in a post. Node colour indicates the proportion of these users who also shared at least one unreliable article, normalised by the total number of users sharing the article ($i.e.$, the node size). Colours range from violet ($0$) to yellow ($1$), with violet indicating a lower prevalence of users who shared reliable content and yellow indicating a higher prevalence of users who exclusively shared reliable content. For visualization purposes, the colour scale is truncated to the interval [$0.1, 0.4$], such that the extreme colours (violet and yellow) correspond to the two boundaries of the truncated scale: violet denotes cases where at least $10\%$ of users sharing the article also shared unreliable content, and yellow denotes cases where at least $40\%$ of users never shared unreliable content}
    \label{fig:doi-co-user-network-reliability}
\end{figure}

Our analysis suggests that the sharing of science by users who also disseminate unreliable sources is not evenly distributed across publications but rather concentrated in noticeable articles (see Fig.~\ref{fig:doi-co-user-network-reliability}). The publication with the highest number of users in this category ($18,224$ users, accounting for $90\%$ of all users who shared this paper) is an article discussing data integrity concerns in Pfizer's COVID-19 vaccine trial \footnote{See \url{https://doi.org/10.1136/bmj.n2635}.}. Similarly, the second most shared publication by users who also shared unreliable sources ($10,524$ users, again representing $90\%$ of all users for this article) presents the claim that ``naturally acquired immunity confers stronger protection against infection and symptomatic disease caused by the Delta variant of SARS-CoV-2''\footnote{See \url{https://doi.org/10.1093/cid/ciac262}.}. In both cases, the central focus of the publications is on vaccine safety and effectivity. Another example involves a study, originally published in 2015, discussing earlier coronaviruses (SARS-CoV and MERS-CoV), which includes an editorial note on the website of the publisher stating that the article ``is being used as the basis for unverified theories that the novel coronavirus causing COVID-19 was engineered''\footnote{See \url{https://doi.org/10.1038/nm.3985}.}. This demonstrates that even earlier literature can become entangled in misinformation narratives later on.

Within this network, we identify the emergence of two main clusters of publications: one cluster consisting of publications predominantly shared by users who also disseminate unreliable sources (mainly violet) and one cluster consisting of users who do not (mainly yellow). One of the most central nodes in this violet cluster corresponds to the previously mentioned publication reporting data integrity issues in Pfizer’s COVID-19 vaccine trial. Moreover, when we shift our focus from unreliable users to users we classified as scientists, a similar pattern emerges (Fig.~\ref{fig:map_sci}). Combined with the pattern in Fig. \ref{fig:doi-co-user-network-reliability}, our results suggest that the publications shared by unreliable users are less likely to be shared by scientists.

%%%%%%%%%%%%%%%%%%%%%%%%%%%%%%%%%%%%%%%%%%%%%%%%%%%%%%%%%%%%%%%%%%%%
%%%%%%%%%%%%%%%%%%%%%%%%%%%%%%%%%%%%%%%%%%%%%%%%%%%%%%%%%%%%%%%%%%%%
%%%%%%%%%%%%%%%%%%%%%%%%%%%%%%%%%%%%%%%%%%%%%%%%%%%%%%%%%%%%%%%%%%%%
%%%%%%%%%%%%%%%%%%%%%%%%%%%%%%%%%%%%%%%%%%%%%%%%%%%%%%%%%%%%%%%%%%%%
%%%%%%%%%%%%%%%%%%%%%%%%%%%%%%%%%%%%%%%%%%%%%%%%%%%%%%%%%%%%%%%%%%%%
\subsection*{Characterising scientists}

As stated previously, we observed that clusters associated with unreliable users tend to have a lower presence of scientists. However, the categories Political and Mainstream Media are included in the definitions of unreliable and reliable, respectively. These categories often display many tweets, but may not be fully representative of scientific content and misinformation. Therefore, in this section, we adopt a narrower perspective: on the one hand, we only consider tweets classified as science, while, on the other hand, we only consider untrustworthy tweets (i.e. containing fake news or conspiracy theories, see the Methods Section for further details). In this way, we are able to explore the interplay between science and more extreme cases of misinformation. 

Notice that, since in this section we only consider posts containing Science and Untrustworthy reliability scores, the analysis is based on a reduced dataset comprising $8,909,163$ posts and $1,599,577$ unique users, among whom $50,233$ are classified as scientists and $1,549,958$ as non-scientists.

\begin{figure}[H]
    \centering
    \includegraphics[width=1\linewidth]{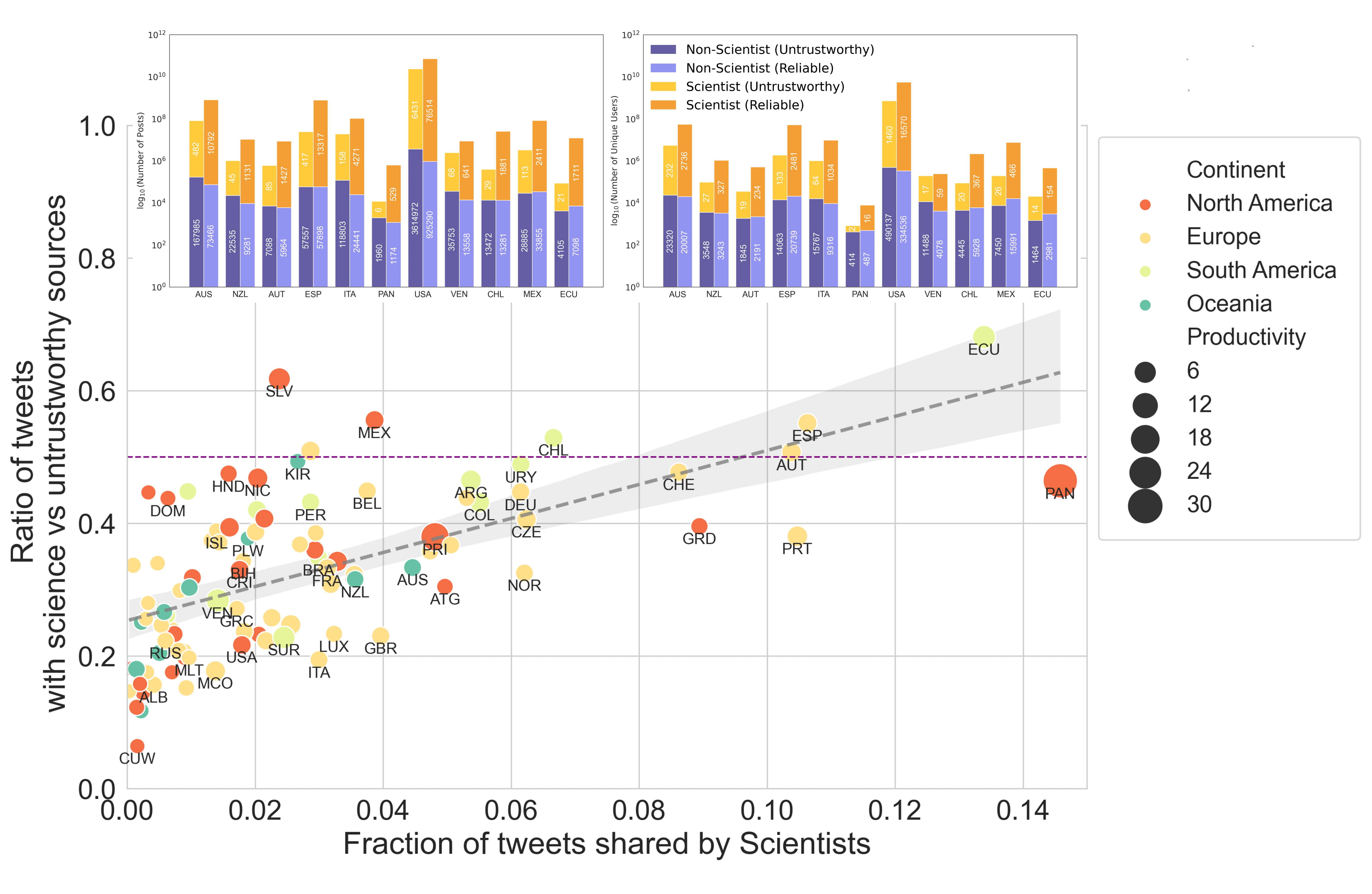}
    \caption{{\bf Interplay between scientific and untrustworthy sources as a function of scientists' Twitter activity}. The x-axis represents the proportion of posts made by scientific users, calculated as the number of posts shared by scientists normalised by the total number of posts in each country. The y-axis shows the ratio of posts containing scientific sources to those containing untrustworthy sources. Data are aggregated by country over the entire period from 2020 to 2023. Countries are colour-coded by continents: North America (which includes Central America and Panama), South America, Europe, and Oceania. The size of the dots corresponds to the average number of tweets per scientist. Bars inside the figure correspond to the total number of posts (left) and the total number of unique users sharing reliable and untrustworthy content for scientists' and non-scientists' users. The y-axis is on a logarithmic scale. We discuss some noteworthy countries in more detail in the main text.}
    \label{fig:correlation}
\end{figure}

Aggregated by country, Fig.~\ref{fig:correlation} shows the correlation between the proportion of scientific content (measured as the number of scientific posts normalised by the total number of scientific and untrustworthy posts) and the level of activity of a scientist on Twitter (the fraction of posts made by scientists with respect to all users). Our results show a clear positive correlation, suggesting that the presence of active scientists on social media is associated with the circulation of scientific results and is simultaneously associated with a lower prominence of untrustworthy content.

From Fig.~\ref{fig:correlation}, we see that a large proportion of countries fall below the line of $0.5$, suggesting that users in these countries tend to share untrustworthy sources more frequently than scientific content. For most countries, the level of scientists active on the social platform during the COVID-19 crisis accounted for less than $10\%$ of the total activity of all users when considering only science and untrustworthy content.
 
Certain countries show a notable engagement in scientific dissemination, such as Spain and Ecuador. As we can see in the bar plots of Fig.~\ref{fig:correlation}, Spain shows a total of $2,507$ scientist users (out of $33,738$ total users), sharing $13,734$ posts (out of $129,189$ total posts), of which $13,317$ contain scientific content and only $417$ include untrustworthy sources, while Ecuador shows a total of $155$ scientist users (out of $4,273$) sharing $1,732$ posts (out of $12,935$ total posts), of which $1,711$ contain scientific content and only $21$ include untrustworthy sources. An interesting case is Panama, where we find only $16$ scientific users out of a total of $825$ users, yet the scientists were very active, sharing $535$ posts containing scientific sources and only $6$ with untrustworthy ones (with a total of $3,669$ posts in the country), suggesting active sharing by scientists during the pandemic. On the other hand, countries such as Mexico and El Salvador have scientists with relatively low activity levels, accounting for less than $5\%$ of the total national activity. However, the number of posts containing scientific sources still exceeds the number of posts with untrustworthy content. This suggests that non-scientist users also contribute to the sharing of scientific information. For example, in Mexico, out of a total of $65,264$ posts, $36,266$ contain scientific sources, while $28,998$ include untrustworthy ones. Surprisingly, the United States shows a relatively low level of scientific activity on the platform compared to general users. Only $2\%$ of tweets are shared by scientists, and the proportion of scientific content is low relative to content from highly untrustworthy sources. This may be explained by the platform's widespread adoption among the general public in the US: a total of $4,623,207$ tweets from the US were collected, of which only $82,945$ were shared by scientists: $1,001,804$ linking to scientific sources and $3,621,403$ to untrustworthy ones. Moreover, only $16,991$ scientists were identified out of a total of $719,535$ users. Although these scientists were relatively active, the general public dominated the content landscape, sharing a total of $3,614,972$ tweets linking to untrustworthy sources and $925,290$ to scientific ones. In the bar plots of Fig.~\ref{fig:correlation}, statistics of other interesting cases can be seen, such as Chile, El Salvador, and Austria, with relatively high levels in both axes compared to countries such as Venezuela, Italy, or New Zealand. Note that this analysis is restricted to only those tweets that contain either scientific content or untrustworthy domains. 

\begin{figure}[H]
    \centering
        \includegraphics[width=1\linewidth]{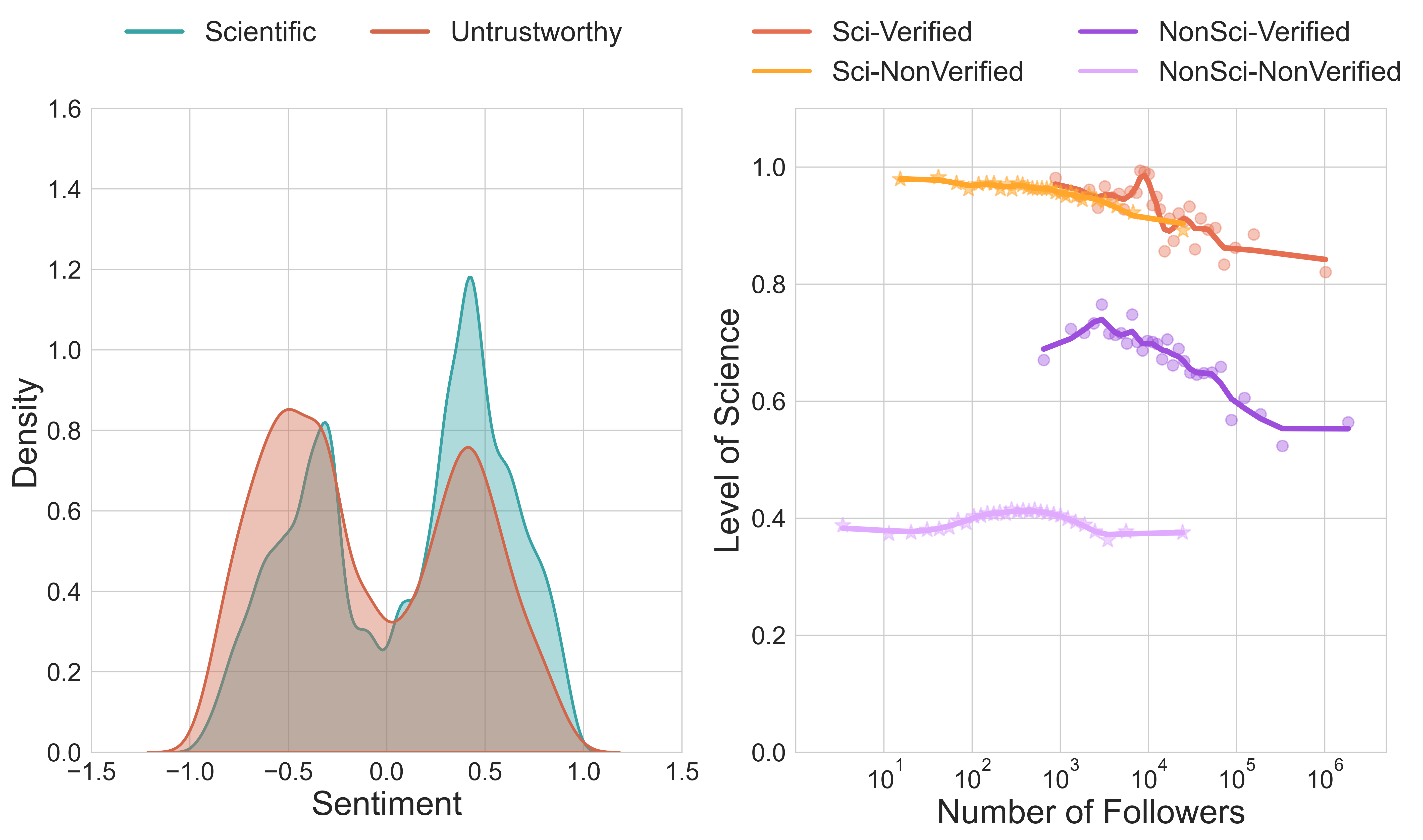}
    \caption{{\bf Statistics on scientific and untrustworthy content}. The left-middle panel displays the density of sentiment of posts containing scientific and untrustworthy content from scientist users. The sentiment is based on \cite{hutto2014vader}, which ranges from $-1$ (most extreme negative) to $1$ (most extreme positive), with $0$ as the neutral point. This metric allows us to assign a single unidimensional measure of emotions expressed in the posts based on the emojis and the text. The right panel represents the level of scientific sources shared by individual users as a function of their number of followers. For each user, we calculate the average proportion of scientific content shared. We consider only posts that contain either scientific sources or untrustworthy sources. We group users in $35$ bins based on their follower count, and we compute the average level of scientific sharing by aggregating across the users within each bin.}
    \label{fig:users-foll}
\end{figure}

We observe that scientists with few followers are typically unverified but exhibit a high level of scientific content sharing in general, with values approaching $1$ (see Fig.~\ref{fig:users-foll}). Counter-intuitively, as the number of followers increases, the proportion of scientific content declines. Furthermore, an overlapping region emerges between verified and non-verified profiles as the number of followers increases. Specifically, there is a gradual drop in the level of scientific sharing among accounts with a large following (exceeding $10^4$ followers). 

Although counter-intuitive, these patterns may be driven by scientists referencing untrustworthy sources with the intention of debunking the source. To better understand behaviour, we analyse the sentiment of scholarly users towards scientific and untrustworthy content, measuring it using the sentiment analysis tool VADER \cite{hutto2014vader}. In Fig.~\ref{fig:users-foll} left-middle panels, we can see that both distributions show two clear peaks, indicating that positive and negative sentiments are present for each type of content. However, for scientific content, the sentiment is more strongly concentrated around a higher positive value, suggesting that scientists generally communicate positively about science. The sharper negative peak for untrustworthy sources may reflect a more consistent critical tone, such as discussing problems or raising concerns. For untrustworthy content, the sentiment distribution shows greater variance and skews more negatively, indicating a wider range of sentiment that combines negativity with neutral or mixed tones.

Finally, for the case of non-scientific users, we can see an opposite trend with respect to the scientific ones. Accounts with a lower number of followers are unverified and tend to share more untrustworthy content with respect to science (see lines below the line of $0.5$), potentially including automated (bot-like) behaviour. Verified users with a high number of followers, including prominent individuals as well as major institutions such as research centres, universities, public health agencies (e.g., WHO, CDC), and international organisations (e.g., the United Nations), tend to share science over highly untrustworthy sources.

%%%%%%%%%%%%%%%%%%%%%%%%%%%%%%%%%%%%%%%%%%%%%%%%%%%%%%%%%%%%%%%%%%%%
%%%%%%%%%%%%%%%%%%%%%%%%%%%%%%%%%%%%%%%%%%%%%%%%%%%%%%%%%%%%%%%%%%%%
%%%%%%%%%%%%%%%%%%%%%%%%%%%%%%%%%%%%%%%%%%%%%%%%%%%%%%%%%%%%%%%%%%%%
%%%%%%%%%%%%%%%%%%%%%%%%%%%%%%%%%%%%%%%%%%%%%%%%%%%%%%%%%%%%%%%%%%%%
%%%%%%%%%%%%%%%%%%%%%%%%%%%%%%%%%%%%%%%%%%%%%%%%%%%%%%%%%%%%%%%%%%%%
%%%%%%%%%%%%%%%%%%%%%%%%%%%%%%%%%%%%%%%%%%%%%%%%%%%%%%%%%%%%%%%%%%%%
%%%%%%%%%%%%%%%%%%%%%%%%%%%%%%%%%%%%%%%%%%%%%%%%%%%%%%%%%%%%%%%%%%%%
%%%%%%%%%%%%%%%%%%%%%%%%%%%%%%%%%%%%%%%%%%%%%%%%%%%%%%%%%%%%%%%%%%%%
%%%%%%%%%%%%%%%%%%%%%%%%%%%%%%%%%%%%%%%%%%%%%%%%%%%%%%%%%%%%%%%%%%%%
%%%%%%%%%%%%%%%%%%%%%%%%%%%%%%%%%%%%%%%%%%%%%%%%%%%%%%%%%%%%%%%%%%%%
%%%%%%%%%%%%%%%%%%%%%%%%%%%%%%%%%%%%%%%%%%%%%%%%%%%%%%%%%%%%%%%%%%%%
%%%%%%%%%%%%%%%%%%%%%%%%%%%%%%%%%%%%%%%%%%%%%%%%%%%%%%%%%%%%%%%%%%%%
%%%%%%%%%%%%%%%%%%%%%%%%%%%%%%%%%%%%%%%%%%%%%%%%%%%%%%%%%%%%%%%%%%%%
%%%%%%%%%%%%%%%%%%%%%%%%%%%%%%%%%%%%%%%%%%%%%%%%%%%%%%%%%%%%%%%%%%%%
%%%%%%%%%%%%%%%%%%%%%%%%%%%%%%%%%%%%%%%%%%%%%%%%%%%%%%%%%%%%%%%%%%%%
\section*{Discussion}

We analyse the interaction between misinformation and science on Twitter (now X) during the COVID-19 pandemic. We find that there is considerable interaction between misinformation and science. Our findings speak to a number of ongoing debates. First, there is an ongoing debate around misinformation and how to address it. Second, there are discussions around the potential misuse of open science, especially in the context of misinformation. We discuss here the implications of our findings for these debates.

As stated in the introduction, defining misinformation in science is challenging and involves clarifying what is considered ``misinformation'' and what is considered ``science''. In our case, some academic publications will be considered ``misinformation'' by other scientists. In some sense, this is reminiscent of discussions around ``junk science'' versus ``sound science'', where not only the substantive results may be discussed, but it is disputed whether something should be considered as ``science''~\cite{jasanoff_contested_1987}. Alternatively, some of the sources marked as unreliable may not necessarily contain misinformation, but may relate to fringe discussions that are not considered legitimate lines of inquiry by the established scientific community. The ongoing discussion in the context of the origins of the coronavirus illustrates these dynamics well, with initial discussions about potential laboratory origins being actively discouraged and being pushed to the fringe, but then subsequently still being investigated by the WHO. The publications used in an unreliable context are not necessarily wrong, and as such do not necessarily directly constitute cases of ``misinformation''~\cite{boumans_fostering_2024}. Rather, it is the interpretation of scientific results and their implications that may be incorrect, which may involve a variety of fallacies. Recent evidence suggests that factually correct information is used to support misinformation narratives~\cite{goel_using_2025}. Other evidence suggests that the sharing of misinformation and a higher general interest in news and politics~\cite{zhou_puzzle_2025} align with our observations. 

There is a tendency to conceive of misinformation as a ``war on science''. However, our results suggest the picture is more nuanced. People who share misinformation also use science to advocate and support their viewpoints. This suggests that science offers a great epistemic reputation upon which to build, even to people who share otherwise unreliable or even untrustworthy sources. That is, there is not necessarily a ``war on science'', and it may not be helpful to frame the problem of misinformation in these terms. In particular, framing the problem as a ``war on science'' may be used to justify closing down some discussions, which may perhaps exacerbate the problem \cite{hardy_effects_2019}. That is, closing down discussions may push potentially legitimate discussions to the fringes, while if such discussions take place in more public venues, a fair and balanced discussion may perhaps diminish potentially problematic interpretations. Of course, this does not negate the need to address and counter known harmful misinformation outright~\cite{krause_infodemic_2022}. But it does question approaching all misinformation as a ``war on science'' \cite{goldenberg_vaccine_2021}, even if some recent developments, most notably in the US, can perhaps be rightfully deemed a ``war on science''.

In a similar vein, our results suggest that ``misinformation'' is not properly addressed by a deficit model~\cite{suldovsky_science_2016}, through increasing engagement with science and increasing science literacy. That is, people who engage with unreliable sources also engage quite clearly with the scientific literature, and their engagement with unreliable sources does not seem to be driven by a lack of interaction with science. Nonetheless, problematic interpretations and common fallacies could perhaps be more actively addressed.

The recent advances in open science, including practices such as open access, open review, and preprinting, may have mixed effects in the interaction with misinformation. On the one hand, greater transparency and availability of results, data, and debates may be beneficial for the trust that the public has in science. It reduces opportunities for selective reporting or publishing and allows others to replicate findings and double-check results. On the other hand, there are potential detrimental effects of such open science practices. Where access was previously mostly limited to members of the scientific community, the open availability of research now allows non-experts to more easily misinterpret and misuse scientific findings. This was, for instance, discussed in the context of genetics research being misused by far-right extremists \cite{carlson_counter_2022}. Anybody may post preprints, possibly containing problematic analyses and interpretations that may be misused. Similarly, with open review, trustworthy results may potentially be discredited through (anonymous) open review platforms by anybody. More research about the effects of open science on public trust in research and on the potential spread of misinformation is needed, but our results hint at some of these potential effects.

We find evidence that academic literature that is available open access or is preprinted is more likely to be used in the context of unreliable sources. This is not to say that only such literature is ``misused''. We also find many publications in reputable journals from reputable publishers that are frequently used by users who also share unreliable sources. From this perspective, our results are mixed, suggesting that although there may potentially be detrimental effects of open science, ``misuses'' are not restricted to open science.

Relatedly, our results clearly show that posts on Twitter about a publication should not necessarily be viewed positively.
That is, a higher Altmetric score, in part informed by the number of tweets, is often viewed positively and may sometimes be suggested to be part of research evaluation. However, a high engagement on Twitter with a paper may reveal particularly problematic interpretations of the findings. One possibility is for services such as Altmetric, PlumX, or Crossref Event Data to play a role in notifying publishers or authors when problematic interpretations emerge on social media.
Such indicators of potential misuse could be based on the approach we developed here.

Moving forward with our analysis, we decided to deepen our understanding by comparing the role of scientists in several countries. Our results suggest that countries where scientists post more show a higher number of tweets with scientific sources. In addition, the average number of posts by scientists in a country plays an important role in promoting a more reliable social media environment. This holds true even in countries with relatively few scientists, such as Panama, where a smaller yet active scientific community can have a positive influence compared to a larger but less active scientific community, such as in the case of the United States. 

The influential prestige of scientists on social media can be approximated by metrics such as follower count and profile verification status. Counter-intuitively, our analysis shows that scientists with a larger following tend to share more untrustworthy content compared to those with a smaller following. One possibility is that scientists with a large following mention untrustworthy sources in order to debunk them. We find that scientists mention untrustworthy content with a more negative sentiment than scientific content, supporting that interpretation. This result needs further investigation to better understand the motivations and contexts behind such sharing behaviour.

Especially in times of crisis, we need scientists to more proactively engage on social media. Our findings suggest that local scientists play a crucial role in delivering trustworthy scientific information, and scientists may, in particular, try to address misinterpretations and misuse of their publications.
 
\subsection*{Limitations}

Although the COVID-19 pandemic presented a unique period in history, especially when it comes to the interaction between science and society, including misinformation, it may not be representative of more general science-society interactions.
That is, much of the evolution of scientific insight during the pandemic was happening in public due to the enormous impact the pandemic had on everyday life across the globe. Moreover, much of the research was publicly available and publicly discussed. Although this offered the public a great view of the kitchen of science, it also highlighted that science is not a static enterprise, with every finding building upon the previous. Instead, it showed that science is an intricate and ever-evolving process, with insights previously considered true being false and vice versa. Possibly, in other situations, there is less engagement with science from users who share unreliable sources.

Moreover, our study relies on reliability classifications based on the domain name of URLs. Although this is not an uncommon approach, and several providers show congruent coding of reliability~\cite{luhring_best_2025}, it does mean that not all information shared from those tweets is necessarily incorrect. As discussed previously, it might also mean that some information was not discussed in more legitimate channels, even if some concerns were justified, or some insights later to be shown true. Although this problem is difficult to overcome, especially at scale, it does raise some questions that could be addressed by more qualitative, in-depth studies considering certain cases in more detail.

This also reveals one weakness of our current approach. Although our study uncovers interaction between misinformation and science, we cannot provide a more nuanced picture of what was discussed and how, if only due to the sheer volume of tweets and publications. Many more detailed case studies could be informed by the dataset that we have created, which could be explored further.

Additionally, we have focused only on Twitter, whereas much of the social media landscape and the spread of misinformation takes place across multiple platforms and formats. Although some of the interactions between science and misinformation may represent general patterns, some of the dynamics may be shaped by the specific affordances of Twitter and could, for instance, play out differently on different platforms.

Finally, although we document some interaction between misinformation and science, and some association with open science, our results do not provide causal effects.
That is, it is not clear whether open science causally affects a publication being more likely to be used in a misinformation context. This would make for an excellent future study.

%%%%%%%%%%%%%%%%%%%%%%%%%%%%%%%%%%%%%%%%%%%%%%%%%%%%%%%%%%%%%%%%%%%%
%%%%%%%%%%%%%%%%%%%%%%%%%%%%%%%%%%%%%%%%%%%%%%%%%%%%%%%%%%%%%%%%%%%%
%%%%%%%%%%%%%%%%%%%%%%%%%%%%%%%%%%%%%%%%%%%%%%%%%%%%%%%%%%%%%%%%%%%%
%%%%%%%%%%%%%%%%%%%%%%%%%%%%%%%%%%%%%%%%%%%%%%%%%%%%%%%%%%%%%%%%%%%%
%%%%%%%%%%%%%%%%%%%%%%%%%%%%%%%%%%%%%%%%%%%%%%%%%%%%%%%%%%%%%%%%%%%%
%%%%%%%%%%%%%%%%%%%%%%%%%%%%%%%%%%%%%%%%%%%%%%%%%%%%%%%%%%%%%%%%%%%%
%%%%%%%%%%%%%%%%%%%%%%%%%%%%%%%%%%%%%%%%%%%%%%%%%%%%%%%%%%%%%%%%%%%%
%%%%%%%%%%%%%%%%%%%%%%%%%%%%%%%%%%%%%%%%%%%%%%%%%%%%%%%%%%%%%%%%%%%%
%%%%%%%%%%%%%%%%%%%%%%%%%%%%%%%%%%%%%%%%%%%%%%%%%%%%%%%%%%%%%%%%%%%%
%%%%%%%%%%%%%%%%%%%%%%%%%%%%%%%%%%%%%%%%%%%%%%%%%%%%%%%%%%%%%%%%%%%%
%%%%%%%%%%%%%%%%%%%%%%%%%%%%%%%%%%%%%%%%%%%%%%%%%%%%%%%%%%%%%%%%%%%%
%%%%%%%%%%%%%%%%%%%%%%%%%%%%%%%%%%%%%%%%%%%%%%%%%%%%%%%%%%%%%%%%%%%%
%%%%%%%%%%%%%%%%%%%%%%%%%%%%%%%%%%%%%%%%%%%%%%%%%%%%%%%%%%%%%%%%%%%%
%%%%%%%%%%%%%%%%%%%%%%%%%%%%%%%%%%%%%%%%%%%%%%%%%%%%%%%%%%%%%%%%%%%%
%%%%%%%%%%%%%%%%%%%%%%%%%%%%%%%%%%%%%%%%%%%%%%%%%%%%%%%%%%%%%%%%%%%%
\section*{Methods}

We built a comprehensive database of messages posted on Twitter (now X) that contains COVID-19 related content. The dataset characterises each tweet, providing information about the statistics (replies, retweets, hashtags used), and the reliability of the content (based on the domain of a URL mentioned in a tweet if there is one). In addition, the dataset contains the user information about users' scientific status, their activity and popularity over time. In addition, when a scientific article is mentioned in a Tweet, the dataset comprises details of the scientific article, such as the discipline, topic, publication status, the journal where it is published, whether it is a preprint, and whether the article is retracted. 

%%%%%%%%%%%%%%%%%%%%%%%%%%%%%%%%%%%%%%%%%%%%%%%%%%%%%%%%%%%%%%%%%%%%
%%%%%%%%%%%%%%%%%%%%%%%%%%%%%%%%%%%%%%%%%%%%%%%%%%%%%%%%%%%%%%%%%%%%
%%%%%%%%%%%%%%%%%%%%%%%%%%%%%%%%%%%%%%%%%%%%%%%%%%%%%%%%%%%%%%%%%%%%
%%%%%%%%%%%%%%%%%%%%%%%%%%%%%%%%%%%%%%%%%%%%%%%%%%%%%%%%%%%%%%%%%%%%
%%%%%%%%%%%%%%%%%%%%%%%%%%%%%%%%%%%%%%%%%%%%%%%%%%%%%%%%%%%%%%%%%%%%
\subsection*{Overview of the Dataset}

We use a large-scale dataset collected by the COVID-19 Infodemic Platform \cite{covid-19Platform}, and explored in detail by \cite{gallotti2020assessing}, from the online social media Twitter (now X) through the application programming interface (API) that provided access to publicly available messages upon specific requests. A set of hashtags and keywords that gained widespread attention (\texttt{coronavirus}, \texttt{ncov}, \texttt{\#Wuhan}, \texttt{covid19}, \texttt{covid-19}, \texttt{sarscov2}, and \texttt{covid}) was used by the authors to collect the tweets; this set includes, for instance, the official names of the virus and disease, as well as the name of the city where the first COVID-19 cases were reported. By mainly using the Streaming API, a random sample of about $1\%$ of the overall (unfiltered) volume of posted messages daily was collected. Regarding the geographic information, approximately $0.8\%$ of the posts were originally geotagged by the users, including the coordinates (latitude-longitude) of the location from which the tweet was posted. To extend this identification, the authors considered the self-defined location of the users to derive the location of the user through a geocoding service, which allows them to geolocate some of the remaining posts. The data may be affected by a possible selection bias of representing well-educated males ($65\%$ of Twitter) between the ages of 18 and 34 ($58\%$ of Twitter users), according to~\cite{statista}. For more details about the methodological choices, see~\cite{gallotti2020assessing}. The messages were in 64 languages from all around the world, but because of data filtering, the largest fraction belongs to English-language sources. 

For the purposes of this research, we include only the subset of tweets containing either a hashtag, URL, or DOI (see the Scientific/COVID-19 subsection) in the textual message. This results in a total of $407,901,738$ messages—including tweets, retweets, replies, and quotes posted on Twitter between January 22, 2020 and March 13, 2023. Our data covers the different stages of the COVID-19 societal debates, including the origin of the virus, the pre-lockdown and lockdown periods, and the development and implementation of the vaccines.

\begin{table}[h!]
    \centering
    \renewcommand{\arraystretch}{1.4}
    \rowcolors{2}{gray!10}{white}
    \begin{tabular}{ l  r }
        \rowcolor{blue!10}
        \textbf{Metric} & \textbf{Count}\\
        Number of tweets & 407,901,738 \\
        Number of users & 27,470,709 \\
        Number of countries & 248 \\
        Number of DOIs & 252,023 \\
    \end{tabular}
    \caption{General statistics on the dataset.}
    \label{table:general-statistics}
\end{table}

%%%%%%%%%%%%%%%%%%%%%%%%%%%%%%%%%%%%%%%%%%%%%%%%%%%%%%%%%%%%%%%%%%%%
%%%%%%%%%%%%%%%%%%%%%%%%%%%%%%%%%%%%%%%%%%%%%%%%%%%%%%%%%%%%%%%%%%%%
%%%%%%%%%%%%%%%%%%%%%%%%%%%%%%%%%%%%%%%%%%%%%%%%%%%%%%%%%%%%%%%%%%%%
%%%%%%%%%%%%%%%%%%%%%%%%%%%%%%%%%%%%%%%%%%%%%%%%%%%%%%%%%%%%%%%%%%%%
%%%%%%%%%%%%%%%%%%%%%%%%%%%%%%%%%%%%%%%%%%%%%%%%%%%%%%%%%%%%%%%%%%%%
\subsection*{Reliability score}

Each tweet containing a URL was assigned a label based on the reliability classification of news media categories as defined by \cite{gallotti2020assessing} and improved by~\cite{bertani2024decoding}. The authors evaluate web domains from various publicly accessible databases covering scientific and journalistic sources to develop a unified classification scheme, as shown in Table~\ref{table:reliability}. They utilised data from Media Bias/FactCheck \cite{MediaBiasFactCheck}, an organisation that provides a large up-to-date database of reliability evaluations of web domains based on reviewing multiple news headline sources. The classification proposed by Media Bias/FactCheck has also been extended with various other open sources, resulting in the identification of $4,417$ domains. Hence, by examining the domain of the URL included in each tweet, we assigned a reliability label. We classified the reliability of tweets based only on the first URL in the tweet; we did not consider multiple URLs.

The domains were classified in the following categories (see Table~\ref{table:reliability}): Science, Mainstream Media, Political, Satire, Clickbait, Fake/Hoax, and Conspiracy/Junk Science. Based on this, we create a coarse-grained categorisation: i) {\bf Reliable}: science and mainstream media, ii) {\bf Unreliable}: satire, clickbait, political, other, fake/hoax, and conspiracy/junk science, iii) {\bf Unknown}: other, shadow, or missing. Tweets without a domain have no reliability categorisation at all. In addition, due to their amplified potential for harm, we subcategorise tweets containing content from Fake/Hoax or Conspiracy/Junk Science domains as {\bf Untrustworthy}, distinguishing them from the broader set of Unreliable sources. Tweets without a domain have no reliability categorisation, and the statistics for each category can be seen in Table \ref{table:reliability-statistics}.

%%%%%%%%%%%%%%%%%%%%%%%%%%%%%%%%%%%%%%%%%%%%%%%%%%%%%%%%%%%%%%%%%%%%
%%%%%%%%%%%%%%%%%%%%%%%%%%%%%%%%%%%%%%%%%%%%%%%%%%%%%%%%%%%%%%%%%%%%
%%%%%%%%%%%%%%%%%%%%%%%%%%%%%%%%%%%%%%%%%%%%%%%%%%%%%%%%%%%%%%%%%%%%
%%%%%%%%%%%%%%%%%%%%%%%%%%%%%%%%%%%%%%%%%%%%%%%%%%%%%%%%%%%%%%%%%%%%
%%%%%%%%%%%%%%%%%%%%%%%%%%%%%%%%%%%%%%%%%%%%%%%%%%%%%%%%%%%%%%%%%%%%
\subsection*{Scientific/COVID-19 data}

We identify whether any of the collected COVID-19 tweets refer to a scientific article on the basis of data from Altmetric\footnote{\url{https://altmetric.com}}, a data analytics platform that tracks the frequency with which scientific articles are mentioned on Twitter and other online platforms. Altmetric also includes data on which publication is mentioned in a tweet, and we gather additional bibliometric data from OpenAlex\footnote{\url{https://openalex.com}}. We calculate the average number of citations for publication sources as the number of citations of the papers published by the source in the period spanning $2020-2022$. We normalise citations on the basis of the field and publication year~\cite{waltman_towards_2011}.

For identifying scientists, we utilized the data gathered by \cite{costas2020large} and \cite{mongeon2023open}. This allows us to identify which Twitter users can be characterised as scientists.  
In addition, we extend our MediaBias/Factcheck classification of the science category (total of $1,891,169$) by incorporating tweets containing DOIs (final total of $7,779,055$). Therefore, any tweet containing a DOI is labelled as science, and this overwrites any prior label.

{\footnotesize
\begin{longtable}{L{3cm} L{3.5cm} L{8.5cm}}
    \rowcolor{blueteal!50}
    \textbf{Category} & \textbf{Type} & \textbf{Description}\\    
    \arrayrulecolor{white}\hline
    \rowcolor{lightblue!70} Science & Reliable & Posts referencing domains whose content has been validated through established scientific scrutiny, or posts containing links that directly or indirectly resolve to Digital Object Identifiers (DOIs).\\ 
    \arrayrulecolor{white}\hline
    \rowcolor{lightblue!30} Mainstream Media & Reliable & Posts containing domains providing content that is generally subjected to professional fact-checking and abides by the rules of media accountability.\\ 
    \arrayrulecolor{white}\hline
    \rowcolor{lightpink!30} Satire & Unreliable & Posts referencing domains providing content that is intentionally and explicitly aiming at providing a distorted representation of events as a form of humour and/or social critique.\\ 
    \arrayrulecolor{white}\hline
    \rowcolor{lightpink!30} Clickbait & Unreliable & Posts referencing domains providing content that generally distorts or intentionally misrepresents information to capture attention.\\ 
    \arrayrulecolor{white}\hline
    \rowcolor{lightpink!30} Political & Unreliable & Posts referencing domains providing content that presents a partisan representation and interpretation of facts to support a political position over rival ones.\\ 
    \arrayrulecolor{white}\hline
    \rowcolor{lightpink!80} Fake/Hoax & Unreliable/Untrustworthy & Posts referencing domains providing manipulative and fabricated content with the purpose of misleading public opinion on socially relevant issues and provoking inflammatory responses.\\ 
    \arrayrulecolor{white}\hline
    \rowcolor{lightpink!80} Conspiracy/Junk Science & Unreliable/Untrustworthy & Posts referencing domains providing systematically manipulative and fabricated content with the purpose of legitimising implausible conceptualisations of facts and knowledge through argumentative methods that coarsely mimic those of scientific reasoning but without any sound logical or factual basis, targeting individuals or social groups as covert instigators or perpetrators of harmful actions.\\ 
    \arrayrulecolor{white}\hline
    \rowcolor{gray!10}Other & Unknown & Posts referencing domains pointing to general content that cannot be easily classified, such as videos on YouTube.\\ 
    \arrayrulecolor{white}\hline
    \rowcolor{gray!10}Shadow & Unknown & Posts referencing domains related to URLs shortening that cannot be classified a priori but would require further URL expansion.\\ 
    \arrayrulecolor{white}\hline
    \rowcolor{gray!10}Missing & Unknown & Posts referencing domains that have not been classified by external experts. \\ 
    \caption{Classification of the reliability score.}
    \label{table:reliability}
\end{longtable}
}

\subsection*{Exposure}

Since tweets were gathered at, or shortly after, the time of posting, we have no indicators of exposure, such as the number of retweets and likes. We therefore rely on an indirect indicator of exposure, namely the number of followers of an account at the time of posting. More formally, we quantify the exposure as the number of followers $K_u(m)$ of the user $u$ who posted a message $m$ at time $t$. The total exposure for a set of tweets is defined as the sum of the number of followers $K_u$. That is, let $S$ be a set of tweets (e.g. reliable tweets, unreliable tweets, tweets in 2020, all tweets), then the exposure $E$ of $S$ is defined as
\begin{equation}
    E(S) = \sum_{m \in S} K_u(m).
\end{equation}

%%%%%%%%%%%%%%%%%%%%%%%%%%%%%%%%%%%%%%%%%%%%%%%%%%%%%%%%%%%%%%%%%%%%
%%%%%%%%%%%%%%%%%%%%%%%%%%%%%%%%%%%%%%%%%%%%%%%%%%%%%%%%%%%%%%%%%%%%
%%%%%%%%%%%%%%%%%%%%%%%%%%%%%%%%%%%%%%%%%%%%%%%%%%%%%%%%%%%%%%%%%%%%
%%%%%%%%%%%%%%%%%%%%%%%%%%%%%%%%%%%%%%%%%%%%%%%%%%%%%%%%%%%%%%%%%%%%
%%%%%%%%%%%%%%%%%%%%%%%%%%%%%%%%%%%%%%%%%%%%%%%%%%%%%%%%%%%%%%%%%%%%
%%%%%%%%%%%%%%%%%%%%%%%%%%%%%%%%%%%%%%%%%%%%%%%%%%%%%%%%%%%%%%%%%%%%
%%%%%%%%%%%%%%%%%%%%%%%%%%%%%%%%%%%%%%%%%%%%%%%%%%%%%%%%%%%%%%%%%%%%
%%%%%%%%%%%%%%%%%%%%%%%%%%%%%%%%%%%%%%%%%%%%%%%%%%%%%%%%%%%%%%%%%%%%
%%%%%%%%%%%%%%%%%%%%%%%%%%%%%%%%%%%%%%%%%%%%%%%%%%%%%%%%%%%%%%%%%%%%
%%%%%%%%%%%%%%%%%%%%%%%%%%%%%%%%%%%%%%%%%%%%%%%%%%%%%%%%%%%%%%%%%%%%
%%%%%%%%%%%%%%%%%%%%%%%%%%%%%%%%%%%%%%%%%%%%%%%%%%%%%%%%%%%%%%%%%%%%
%%%%%%%%%%%%%%%%%%%%%%%%%%%%%%%%%%%%%%%%%%%%%%%%%%%%%%%%%%%%%%%%%%%%
%%%%%%%%%%%%%%%%%%%%%%%%%%%%%%%%%%%%%%%%%%%%%%%%%%%%%%%%%%%%%%%%%%%%
%%%%%%%%%%%%%%%%%%%%%%%%%%%%%%%%%%%%%%%%%%%%%%%%%%%%%%%%%%%%%%%%%%%%
%%%%%%%%%%%%%%%%%%%%%%%%%%%%%%%%%%%%%%%%%%%%%%%%%%%%%%%%%%%%%%%%%%%%    
\section*{Acknowledgements}

We gratefully acknowledge the contribution of Manlio De Domenico, who led the COVID-19 Infodemic Observatory project that enabled the collection of the dataset analysed in this work. This project is entirely funded by the European Media and Information Fund,`Understanding Misinformation and Science in Societal Debates' (UnMiSSeD) project.

\section*{Data availability}

Data at the level of DOIs is available in Zenodo~\cite{alvarez_zuzek_mapping_2025}. Data at the level of individuals tweets cannot be made available due to licence restrictions.

\newpage
\renewcommand\thefigure{S\arabic{figure}}
\setcounter{figure}{0}
\renewcommand\thetable{S\arabic{table}}
\setcounter{table}{0}
%%%%%%%%%%%%%%%%%%%%%%%%%%%%%%%%%%%%%%%%%%%%%%%%%%%%%%%%%%%%%%%%%%%%
%%%%%%%%%%%%%%%%%%%%%%%%%%%%%%%%%%%%%%%%%%%%%%%%%%%%%%%%%%%%%%%%%%%%
%%%%%%%%%%%%%%%%%%%%%%%%%%%%%%%%%%%%%%%%%%%%%%%%%%%%%%%%%%%%%%%%%%%%
%%%%%%%%%%%%%%%%%%%%%%%%%%%%%%%%%%%%%%%%%%%%%%%%%%%%%%%%%%%%%%%%%%%%
%%%%%%%%%%%%%%%%%%%%%%%%%%%%%%%%%%%%%%%%%%%%%%%%%%%%%%%%%%%%%%%%%%%%
%%%%%%%%%%%%%%%%%%%%%%%%%%%%%%%%%%%%%%%%%%%%%%%%%%%%%%%%%%%%%%%%%%%%
%%%%%%%%%%%%%%%%%%%%%%%%%%%%%%%%%%%%%%%%%%%%%%%%%%%%%%%%%%%%%%%%%%%%
%%%%%%%%%%%%%%%%%%%%%%%%%%%%%%%%%%%%%%%%%%%%%%%%%%%%%%%%%%%%%%%%%%%%
%%%%%%%%%%%%%%%%%%%%%%%%%%%%%%%%%%%%%%%%%%%%%%%%%%%%%%%%%%%%%%%%%%%%
%%%%%%%%%%%%%%%%%%%%%%%%%%%%%%%%%%%%%%%%%%%%%%%%%%%%%%%%%%%%%%%%%%%%
%%%%%%%%%%%%%%%%%%%%%%%%%%%%%%%%%%%%%%%%%%%%%%%%%%%%%%%%%%%%%%%%%%%%
%%%%%%%%%%%%%%%%%%%%%%%%%%%%%%%%%%%%%%%%%%%%%%%%%%%%%%%%%%%%%%%%%%%%
%%%%%%%%%%%%%%%%%%%%%%%%%%%%%%%%%%%%%%%%%%%%%%%%%%%%%%%%%%%%%%%%%%%%
%%%%%%%%%%%%%%%%%%%%%%%%%%%%%%%%%%%%%%%%%%%%%%%%%%%%%%%%%%%%%%%%%%%%
%%%%%%%%%%%%%%%%%%%%%%%%%%%%%%%%%%%%%%%%%%%%%%%%%%%%%%%%%%%%%%%%%%%% 
\section*{Appendix}

\subsection*{General statistics for tweets}

In this section, we present more detailed statistics of the characteristics of our dataset. 

{\footnotesize
\begin{longtable}{l c c c c c }
    \rowcolor{blueteal!50}
    \textbf{Category} & \textbf{\% of Tweets} & \textbf{\% of Users} & \textbf{\% of Exposure} & \textbf{Avg Followers} & \textbf{Median Followers} \\
    \arrayrulecolor{white}\hline
    All Tweets & \cellcolor{green4}100.00 & \cellcolor{green4}100.00 & \cellcolor{green4}100.00 & \cellcolor{green2}34,586 & \cellcolor{green1}676 \\
    \arrayrulecolor{white}\hline
    No reliability score & \cellcolor{green3}57.91 & \cellcolor{green4}80.08 & \cellcolor{green2}36.43 & \cellcolor{green1}21,761 & \cellcolor{green1}606 \\
    \arrayrulecolor{white}\hline   
    \rowcolor{gray!20}Shadow & \cellcolor{green1}7.96 & \cellcolor{green2}20.54 & \cellcolor{green3}24.44 & \cellcolor{green4}106,239 & \cellcolor{green3}855 \\
    \arrayrulecolor{white}\hline
    \rowcolor{gray!20}Other & \cellcolor{green0}1.54 & \cellcolor{green1}7.65 & \cellcolor{green1}0.83 & \cellcolor{green1}18,608 & \cellcolor{green1}616 \\
    \arrayrulecolor{white}\hline
    \rowcolor{gray!20}Missing & \cellcolor{green2}12.10 & \cellcolor{green3}27.98 & \cellcolor{green2}14.15 & \cellcolor{green2}40,449 & \cellcolor{green2}802 \\
    \arrayrulecolor{white}\hline
    \rowcolor{gray!60}(total) Unknown reliability & \cellcolor{green3}21.59 & \cellcolor{green3}41.20 & \cellcolor{green3}39.42 & \cellcolor{green3}63,138 & \cellcolor{green2}807 \\
    \arrayrulecolor{white}\hline
    \rowcolor{lightblue!30}Mainstream Media & \cellcolor{green2}12.92 & \cellcolor{green3}26.45 & \cellcolor{green2}18.16 & \cellcolor{green3}48,602 & \cellcolor{green2}784 \\
    \arrayrulecolor{white}\hline
    \rowcolor{lightblue!60}Science & \cellcolor{green0}1.91 & \cellcolor{green0}5.89 & \cellcolor{green0}0.38 & \cellcolor{green0}6,837 & \cellcolor{green1}700 \\
    \arrayrulecolor{white}\hline
    \rowcolor{lightblue!90}(total) Reliable & \cellcolor{green2}14.83 & \cellcolor{green3}28.36 & \cellcolor{green2}18.53 & \cellcolor{green3}43,231 & \cellcolor{green2}772 \\
    \arrayrulecolor{white}\hline
    \rowcolor{lightpink!30}Political & \cellcolor{green1}3.68 & \cellcolor{green2}10.85 & \cellcolor{green1}4.89 & \cellcolor{green3}45,953 & \cellcolor{green2}780 \\
    \arrayrulecolor{white}\hline
    \rowcolor{lightpink!30}Clickbait & \cellcolor{green0}0.34 & \cellcolor{green1}2.24 & \cellcolor{green0}0.33 & \cellcolor{green2}34,074 & \cellcolor{green0}552 \\
    \arrayrulecolor{white}\hline
    \rowcolor{lightpink!30}Satire & \cellcolor{green0}0.05 & \cellcolor{green0}0.42 & \cellcolor{green0}0.01 & \cellcolor{green0}4,406 & \cellcolor{green0}482 \\
    \arrayrulecolor{white}\hline
    \rowcolor{lightpink!80}Fake \& Hoax & \cellcolor{green0}1.19 & \cellcolor{green1}3.23 & \cellcolor{green0}0.34 & \cellcolor{green0}9,857 & \cellcolor{green3}933 \\
    \arrayrulecolor{white}\hline
    \rowcolor{lightpink!80}Conspiracy \& Junk science & \cellcolor{green0}0.41 & \cellcolor{green0}1.33 & \cellcolor{green0}0.04 & \cellcolor{green0}3,697 & \cellcolor{green1}717 \\
    \arrayrulecolor{white}\hline
    \rowcolor{darkpink!60}Untrustworthy & \cellcolor{green0}1.60 & \cellcolor{green1}3.73 & \cellcolor{green0}0.38 & \cellcolor{green0}8,274 & \cellcolor{green3}867 \\
    \arrayrulecolor{white}\hline
    \rowcolor{darkpink!80}(total) Unreliable & \cellcolor{green1}5.67 & \cellcolor{green2}13.81 & \cellcolor{green1}5.61 & \cellcolor{green2}34,243 & \cellcolor{green2}782 \\

    \caption{{\bf Summary statistics of the various reliability categories.}}
    \label{table:reliability-statistics}
\end{longtable}
}

%%%%%%%%%%%%%%%%%%%%%%%%%%%%%%%%%%%%%%%%%%%%%%%%%%%%%%%%%%%%%%%%%%%%
%%%%%%%%%%%%%%%%%%%%%%%%%%%%%%%%%%%%%%%%%%%%%%%%%%%%%%%%%%%%%%%%%%%%
%%%%%%%%%%%%%%%%%%%%%%%%%%%%%%%%%%%%%%%%%%%%%%%%%%%%%%%%%%%%%%%%%%%%
%%%%%%%%%%%%%%%%%%%%%%%%%%%%%%%%%%%%%%%%%%%%%%%%%%%%%%%%%%%%%%%%%%%%
%%%%%%%%%%%%%%%%%%%%%%%%%%%%%%%%%%%%%%%%%%%%%%%%%%%%%%%%%%%%%%%%%%%%
\subsection*{Timeline}

\begin{figure}[H]
    \centering
    \includegraphics[width=1\linewidth]{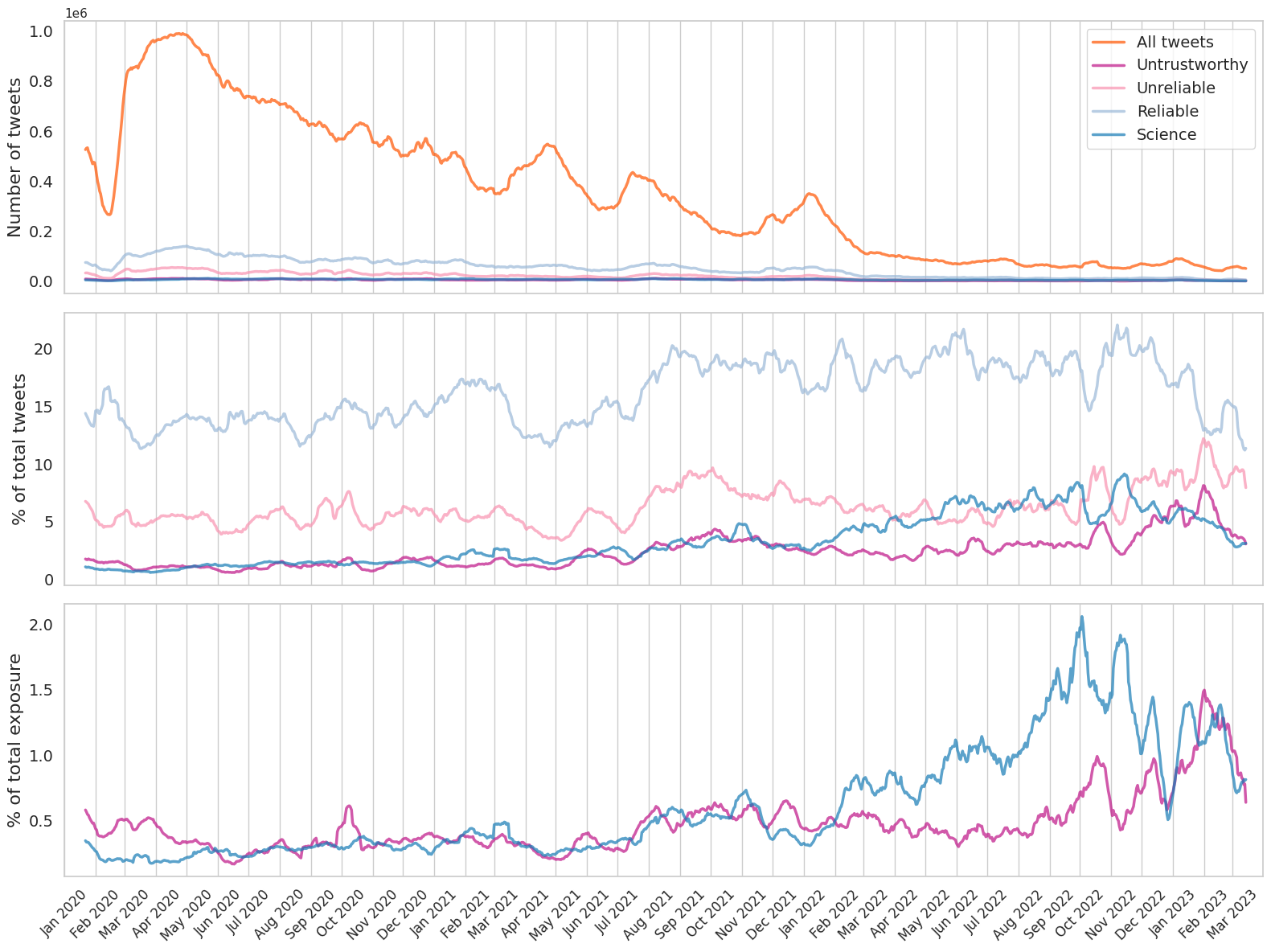}
    \caption{{\bf Tweet statistics over time}. Our dataset spans from January 22, 2020, to March 18, 2023. Panel (a) shows the total number of tweets over time, including: all tweets in orange, tweets containing untrustworthy in dark pink, unreliable sources in pink (containing also untrustworthy), reliable in light blue (containing also science), and science in dark blue. While (b) provides a focused view of the proportion of tweets, excluding the all-tweets curve for better clarity. Panel (d) is the proportion of exposure only for the extreme cases of reliability, untrustworthy, and science.}
    \label{fig:timeline}
\end{figure}

The data collection of the COVID-19 infodemics observatory \cite{gallotti2020assessing} started January 22, 2020, using the Twitter Filter API v1 and a set of keywords describing the COVID-19 epidemics from a medical perspective (see Methods Section). Out of the already reduced sample of the Twitter API, the platform filters only about 50\% of users that can be associated with a specific country. In the upper panel of Fig.~\ref{fig:timeline}, we see how the number of Tweets collected grew rapidly during the first weeks of the collection, reaching its peak at the end of April. 

%%%%%%%%%%%%%%%%%%%%%%%%%%%%%%%%%%%%%%%%%%%%%%%%%%%%%%%%%%%%%%%%%%%%
%%%%%%%%%%%%%%%%%%%%%%%%%%%%%%%%%%%%%%%%%%%%%%%%%%%%%%%%%%%%%%%%%%%%
%%%%%%%%%%%%%%%%%%%%%%%%%%%%%%%%%%%%%%%%%%%%%%%%%%%%%%%%%%%%%%%%%%%%
%%%%%%%%%%%%%%%%%%%%%%%%%%%%%%%%%%%%%%%%%%%%%%%%%%%%%%%%%%%%%%%%%%%%
%%%%%%%%%%%%%%%%%%%%%%%%%%%%%%%%%%%%%%%%%%%%%%%%%%%%%%%%%%%%%%%%%%%%
\subsection*{General statistics per user}

Our final dataset has a total of $27,294,925$ unique users, where only $194,668$ are classified as scientists, representing only $0.71\%$ of the users. However, scientists are more active on average compared with non-scientists. Scientists are much more likely to share tweets with DOIs, representing $14.7\%$ of their tweets, compared to only $1.3\%$ with non-scientists. Although scientists share fewer tweets with unreliable sources, still $0.2\%$ of their tweets represent fake news/hoax, and $0.1\%$ conspiracy theories and junk science.

\newpage
{\footnotesize
\begin{longtable}{L{1.8cm} L{1cm} L{1cm} L{1cm} L{1cm} L{1cm} L{1cm} L{1cm} L{1cm} L{1cm} L{1cm}}
    \rowcolor{blueteal!50}
        \textbf{} & \textbf{\scriptsize{SCI.\%}} & \textbf{\scriptsize{MSM.\%}} & \textbf{\scriptsize{SAT.\%}} & \textbf{\scriptsize{CLI.\%}} & \textbf{\scriptsize{POLI.\%}} & \textbf{\scriptsize{FAKE.\%}} & \textbf{\scriptsize{CON.\%}} & \textbf{\scriptsize{OTH.\%}} & \textbf{\scriptsize{SHAD.\%}} & \textbf{\scriptsize{MISS.\%}} \\

        \arrayrulecolor{white}\hline
        \cellcolor{gray!20}Other & \cellcolor{green2}23.12 & \cellcolor{green3}53.63 & \cellcolor{green0}2.35 & \cellcolor{green1}7.48 & \cellcolor{green2}34.97 & \cellcolor{green2}17.03 & \cellcolor{green1}8.88 & -- & \cellcolor{green3}43.24 & \cellcolor{green3}56.43 \\
        \arrayrulecolor{white}\hline
        \cellcolor{gray!20}Shadow & \cellcolor{green1}14.14 & \cellcolor{green3}50.28 & \cellcolor{green0}1.43 & \cellcolor{green1}5.71 & \cellcolor{green2}28.55 & \cellcolor{green1}9.77 & \cellcolor{green0}4.46 & \cellcolor{green2}16.10 & -- & \cellcolor{green3}49.48 \\
        \arrayrulecolor{white}\hline
        \cellcolor{gray!20}Missing & \cellcolor{green1}13.17 & \cellcolor{green3}44.64 & \cellcolor{green0}1.05 & \cellcolor{green1}4.25 & \cellcolor{green2}23.15 & \cellcolor{green1}7.96 & \cellcolor{green0}3.58 & \cellcolor{green2}15.42 & \cellcolor{green3}36.32 & -- \\
        \arrayrulecolor{white}\hline
        \cellcolor{lightblue!30}Mainstream Media & \cellcolor{green1}15.03 & -- & \cellcolor{green0}1.29 & \cellcolor{green0}5.11 & \cellcolor{green2}27.18 & \cellcolor{green1}9.27 & \cellcolor{green0}4.13 & \cellcolor{green1}15.51 & \cellcolor{green2}39.05 & \cellcolor{green3}47.23 \\
        \arrayrulecolor{white}\hline
        \cellcolor{lightblue!60}Science & -- & \cellcolor{green4}67.52 & \cellcolor{green0}3.26 & \cellcolor{green1}8.95 & \cellcolor{green3}40.81 & \cellcolor{green2}20.11 & \cellcolor{green1}10.98 & \cellcolor{green2}30.03 & \cellcolor{green3}49.32 & \cellcolor{green4}62.59 \\
        \arrayrulecolor{white}\hline
        \cellcolor{lightpink!30}Satire & \cellcolor{green3}45.28 & \cellcolor{green4}80.27 & -- & \cellcolor{green2}21.43 & \cellcolor{green4}65.00 & \cellcolor{green3}46.65 & \cellcolor{green2}32.66 & \cellcolor{green2}42.36 & \cellcolor{green4}69.60 & \cellcolor{green4}69.43 \\
        \arrayrulecolor{white}\hline
        \cellcolor{lightpink!30}Clickbait & \cellcolor{green2}23.53 & \cellcolor{green4}60.41 & \cellcolor{green0}4.05 & -- & \cellcolor{green3}38.24 & \cellcolor{green2}26.60 & \cellcolor{green1}14.36 & \cellcolor{green1}25.56 & \cellcolor{green3}52.36 & \cellcolor{green3}53.08 \\
        \arrayrulecolor{white}\hline
        \cellcolor{lightpink!30}Political & \cellcolor{green2}22.14 & \cellcolor{green4}66.21 & \cellcolor{green0}2.54 & \cellcolor{green1}7.89 & -- & \cellcolor{green2}18.19 & \cellcolor{green1}8.44 & \cellcolor{green2}24.65 & \cellcolor{green4}54.02 & \cellcolor{green4}59.67 \\
        \arrayrulecolor{white}\hline
        \cellcolor{lightpink!80}Fake \& Hoax & \cellcolor{green3}36.61 & \cellcolor{green4}75.84 & \cellcolor{green1}6.11 & \cellcolor{green2}18.42 & \cellcolor{green4}61.06 & -- & \cellcolor{green2}25.77 & \cellcolor{green2}40.29 & \cellcolor{green3}62.04 & \cellcolor{green4}68.89 \\
        \arrayrulecolor{white}\hline
        \cellcolor{lightpink!80}Conspiracy \& junk science & \cellcolor{green4}48.48 & \cellcolor{green4}81.83 & \cellcolor{green1}10.37 & \cellcolor{green2}24.10 & \cellcolor{green4}68.71 & \cellcolor{green4}62.46 & -- & \cellcolor{green4}50.91 & \cellcolor{green4}68.61 & \cellcolor{green4}75.18 \\
    \caption{{\bf Pairwise overlap percentages between reliability categories}. Each row shows the percentage of users who shared content (i.e. at least one tweet) in that row who also share content (i.e. at least one tweet) in the category in the column. Note that percentages do not sum to $100\%$ because users may share content from multiple categories.}
    \label{table:pairwise-overlap}
\end{longtable}
}

%%%%%%%%%%%%%%%%%%%%%%%%%%%%%%%%%%%%%%%%%%%%%%%%%%%%%%%%%%%%%%%%%%%%
%%%%%%%%%%%%%%%%%%%%%%%%%%%%%%%%%%%%%%%%%%%%%%%%%%%%%%%%%%%%%%%%%%%%
%%%%%%%%%%%%%%%%%%%%%%%%%%%%%%%%%%%%%%%%%%%%%%%%%%%%%%%%%%%%%%%%%%%%
%%%%%%%%%%%%%%%%%%%%%%%%%%%%%%%%%%%%%%%%%%%%%%%%%%%%%%%%%%%%%%%%%%%%
%%%%%%%%%%%%%%%%%%%%%%%%%%%%%%%%%%%%%%%%%%%%%%%%%%%%%%%%%%%%%%%%%%%%
\subsection*{Correlations per users}

\begin{figure}[H]
    \centering
    \includegraphics[width=0.45\linewidth]{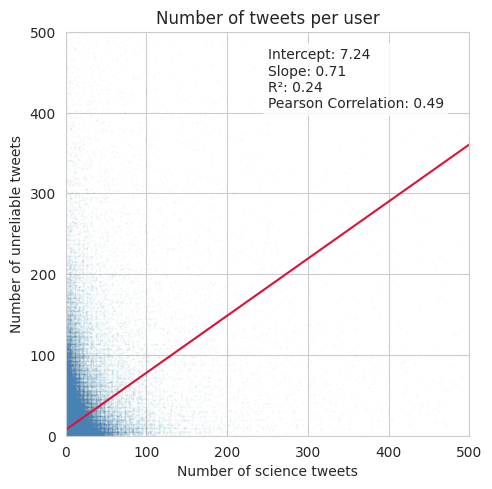}
    \includegraphics[width=0.45\linewidth]{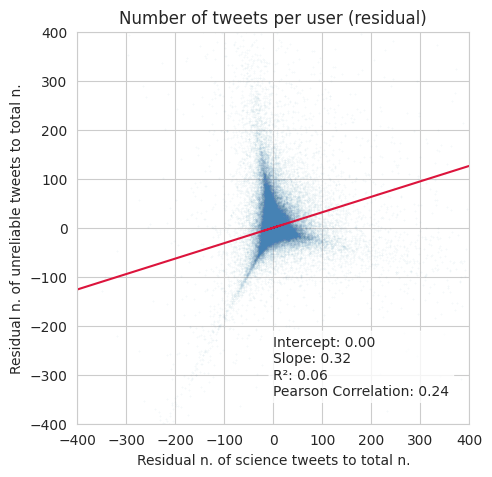}
    \includegraphics[width=0.45\linewidth]{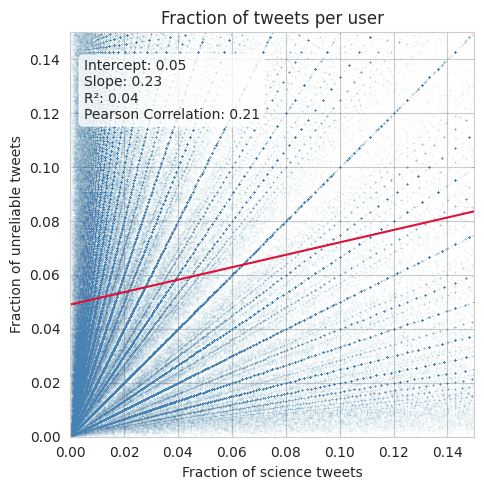}
    \caption{(a) User-wise correlation between the number of unreliable tweets and the number of DOI tweets. (b) User-wise partial correlation between the residual number of unreliable tweets and the residual number of DOI tweets after regressing on the total number of tweets. (c) User-wise correlation between the fraction of unreliable tweets and the fraction of DOI tweets.}
    \label{fig:correlation_per_user}
\end{figure}

%%%%%%%%%%%%%%%%%%%%%%%%%%%%%%%%%%%%%%%%%%%%%%%%%%%%%%%%%%%%%%%%%%%%
%%%%%%%%%%%%%%%%%%%%%%%%%%%%%%%%%%%%%%%%%%%%%%%%%%%%%%%%%%%%%%%%%%%%
%%%%%%%%%%%%%%%%%%%%%%%%%%%%%%%%%%%%%%%%%%%%%%%%%%%%%%%%%%%%%%%%%%%%
%%%%%%%%%%%%%%%%%%%%%%%%%%%%%%%%%%%%%%%%%%%%%%%%%%%%%%%%%%%%%%%%%%%%
%%%%%%%%%%%%%%%%%%%%%%%%%%%%%%%%%%%%%%%%%%%%%%%%%%%%%%%%%%%%%%%%%%%%
\subsection*{Topic landscape}

{\footnotesize
\begin{longtable}{l p{5cm} r r}
    \rowcolor{blue!10}
    \textbf{Visualization label} & \textbf{Long name} & \textbf{DOIs} & \textbf{\% unreliably used DOIs} \\
    COVID-19 & Coronavirus Disease 2019 Research & 2274 & \colorPercentCell{56.60} \\
    \rowcolor{gray!10}COVID-19 & Coronavirus Disease 2019 & 1298 & \colorPercentCell{33.36} \\
    COVID-19 Neurology & Neurological Manifestations of COVID-19 Infection & 623 & \colorPercentCell{68.54} \\
    \rowcolor{gray!10}COVID-19 Modeling & Modeling the Dynamics of COVID-19 Pandemic & 583 & \colorPercentCell{51.46} \\
    Airborne Transmission & Airborne Transmission of Respiratory Viruses & 409 & \colorPercentCell{63.33} \\
    \rowcolor{gray!10}COVID-19 Diagnostics & Diagnostic Methods for COVID-19 Detection & 210 & \colorPercentCell{50.00} \\
    Vaccine Hesitancy & Factors Affecting Vaccine Hesitancy and Acceptance & 170 & \colorPercentCell{45.88} \\
    \rowcolor{gray!10}COVID-19 in Pregnancy & Impact of COVID-19 Infection on Pregnancy Outcomes & 160 & \colorPercentCell{32.50} \\
    Cancer and COVID-19 & Impact of COVID-19 on Cancer Patients and Care & 156 & \colorPercentCell{45.51} \\
    \rowcolor{gray!10}Mental Health & Impact of COVID-19 on Mental Health & 147 & \colorPercentCell{23.13} \\
    Vitamin D & Vitamin D and Health Outcomes & 115 & \colorPercentCell{75.65} \\
    \rowcolor{gray!10}Parasitic Diseases & Parasitic Diseases and Treatment Strategies & 96 & \colorPercentCell{76.04} \\
    Kawasaki Disease & Diagnosis and Management of Kawasaki Disease & 55 & \colorPercentCell{54.55} \\
    \rowcolor{gray!10}Influenza & Influenza Virus Research and Epidemiology & 41 & \colorPercentCell{60.98} \\
    Mechanical Ventilation & Mechanical Ventilation in Respiratory Failure and ARDS & 40 & \colorPercentCell{0.00} \\
    \rowcolor{gray!10}Heparin-Induced Thrombocytopenia & Heparin-Induced Thrombocytopenia and Thrombosis & 40 & \colorPercentCell{77.50} \\
    Healthcare Workforce & Impact of Healthcare Workforce on Public Health & 39 & \colorPercentCell{82.05} \\
    \rowcolor{gray!10}Misinformation & The Spread of Misinformation Online & 36 & \colorPercentCell{36.11} \\
    Respiratory Viral Infections & Epidemiology and Pathogenesis of Respiratory Viral Infections & 32 & \colorPercentCell{62.50} \\
    \rowcolor{gray!10}Olfactory Dysfunction & Olfactory Dysfunction in Health and Disease & 31 & \colorPercentCell{29.03} \\
    Gender Bias & Gender Bias in Academic Medicine and Science & 28 & \colorPercentCell{3.57} \\
    \rowcolor{gray!10}Nursing Home Care & Quality and Practices in Nursing Home Care & 25 & \colorPercentCell{40.00} \\
    Trained Immunity & Trained Immunity in Health and Disease & 25 & \colorPercentCell{32.00} \\
    \caption{Leiden ranking microclusters with 20 or more DOIs.
    The percentage of unreliably used publications refers to publications that are more often mentioned by users who also share unreliable sources than the median.} 
    \label{fig:topics-dois}
\end{longtable}
}

{\footnotesize
\begin{longtable}{L{8cm} r c }
    \rowcolor{blue!10}
    \textbf{Source} & \textbf{DOIs} & \textbf{\% unreliably used DOIs}\\    
    medRxiv & 763 & \percentColorCell{63.17} \\
    \rowcolor{gray!10} Nature & 683 & \percentColorCell{38.07} \\
    BMJ & 457 & \percentColorCell{54.49} \\
    \rowcolor{gray!10} JAMA & 319 & \percentColorCell{26.33} \\
    The New England Journal of Medicine & 315 & \percentColorCell{31.11} \\
    \rowcolor{gray!10} The Lancet & 307 & \percentColorCell{35.83} \\
    Science & 289 & \percentColorCell{25.95} \\
    \rowcolor{gray!10} Morbidity and Mortality Weekly Report & 261 & \percentColorCell{65.13} \\
    bioRxiv & 210 & \percentColorCell{57.14} \\
    Nature Medicine & 129 & \percentColorCell{27.91} \\
    \rowcolor{gray!10} JAMA network open & 121 & \percentColorCell{64.46} \\
    Lancet Infectious Diseases & 106 & \percentColorCell{57.55} \\
    \rowcolor{gray!10} Proceedings of the National Academy of Sciences (USA) & 102 & \percentColorCell{54.90} \\
    Cell & 89 & \percentColorCell{21.35} \\
    \rowcolor{gray!10} Clinical Infectious Diseases & 88 & \percentColorCell{69.32} \\
    The Lancet Respiratory Medicine & 74 & \percentColorCell{24.32} \\
    \rowcolor{gray!10} Nature Communications & 70 & \percentColorCell{50.00} \\
    JAMA Internal Medicine & 68 & \percentColorCell{42.65} \\
    \rowcolor{gray!10} Annals of Internal Medicine & 64 & \percentColorCell{56.25} \\
    Emerging Infectious Diseases & 63 & \percentColorCell{80.95} \\
    \rowcolor{gray!10} Research Square & 59 & \percentColorCell{74.58} \\
    Scientific Reports & 58 & \percentColorCell{77.59} \\
    \rowcolor{gray!10} Science immunology & 45 & \percentColorCell{24.44} \\
    Intensive Care Medicine & 44 & \percentColorCell{2.27} \\
    \rowcolor{gray!10} Science Translational Medicine & 43 & \percentColorCell{48.84} \\
    Science Advances & 40 & \percentColorCell{27.50} \\
    \rowcolor{gray!10} The Lancet Child \& Adolescent Health & 37 & \percentColorCell{62.16} \\
    JAMA Pediatrics & 36 & \percentColorCell{52.78} \\
    \rowcolor{gray!10} International Journal of Infectious Diseases & 35 & \percentColorCell{85.71} \\
    PLOS ONE & 35 & \percentColorCell{68.57} \\
    \rowcolor{gray!10} EClinicalMedicine & 35 & \percentColorCell{77.14} \\
    Nature Reviews Immunology & 34 & \percentColorCell{20.59} \\
    \rowcolor{gray!10} Cureus & 33 & \percentColorCell{93.94} \\
    Immunity & 31 & \percentColorCell{35.48} \\
    \rowcolor{gray!10} Nature Biotechnology & 30 & \percentColorCell{13.33} \\
    Eurosurveillance & 29 & \percentColorCell{65.52} \\
    \rowcolor{gray!10} The Lancet Public Health & 27 & \percentColorCell{22.22} \\
    Journal of Infection & 27 & \percentColorCell{81.48} \\
    \rowcolor{gray!10} Circulation & 24 & \percentColorCell{41.67} \\
    The Cochrane library & 23 & \percentColorCell{13.04} \\
    \rowcolor{gray!10} The Lancet Global Health & 23 & \percentColorCell{39.13} \\
    Canadian Medical Association Journal & 21 & \percentColorCell{61.90} \\
    \rowcolor{gray!10} Frontiers in Immunology & 21 & \percentColorCell{85.71} \\
    The Lancet Diabetes \& Endocrinology & 21 & \percentColorCell{33.33} \\
    \rowcolor{gray!10} Clinical Microbiology and Infection & 21 & \percentColorCell{52.38} \\
    The Lancet Rheumatology & 20 & \percentColorCell{35.00} \\
    \rowcolor{gray!10} Nature Immunology & 20 & \percentColorCell{60.00} \\
    \caption{Sources with 20 or more DOIs.
    The percentage of unreliably used publications refers to publications that are more often mentioned by users who also share unreliable sources than the median.}
    \label{table:sources-dois}
\end{longtable}
}

%%%%%%%%%%%%%%%%%%%%%%%%%%%%%%%%%%%%%%%%%%%%%%%%%%%%%%%%%%%%%%%%%%%%
%%%%%%%%%%%%%%%%%%%%%%%%%%%%%%%%%%%%%%%%%%%%%%%%%%%%%%%%%%%%%%%%%%%%
%%%%%%%%%%%%%%%%%%%%%%%%%%%%%%%%%%%%%%%%%%%%%%%%%%%%%%%%%%%%%%%%%%%%
%%%%%%%%%%%%%%%%%%%%%%%%%%%%%%%%%%%%%%%%%%%%%%%%%%%%%%%%%%%%%%%%%%%%
%%%%%%%%%%%%%%%%%%%%%%%%%%%%%%%%%%%%%%%%%%%%%%%%%%%%%%%%%%%%%%%%%%%%
\subsection*{Characterising publications}

\begin{figure}[H]
    \centering
    \includegraphics[width=1\linewidth]{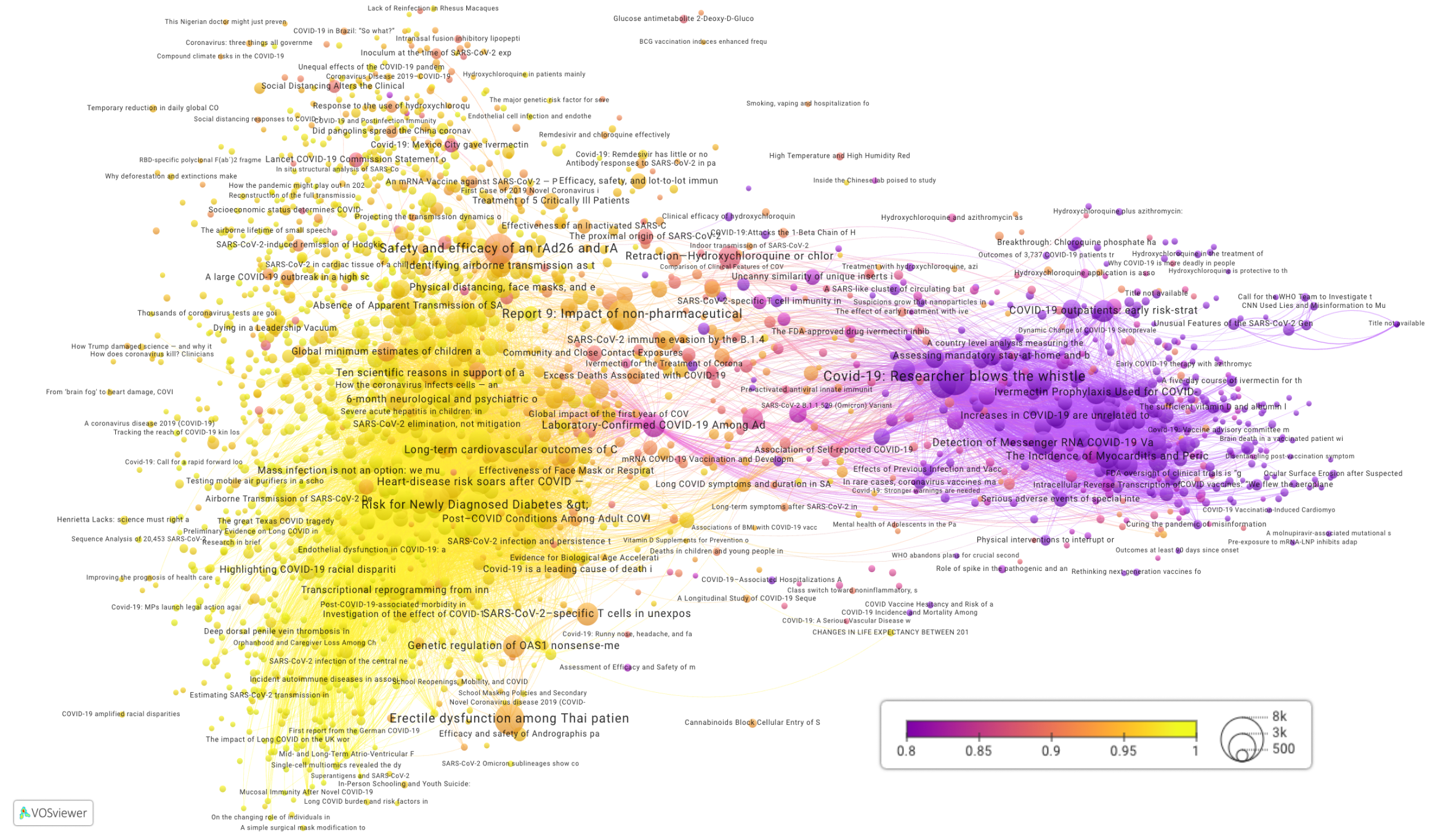}
    \caption{{\bf Network of publications co-shared by users according to reliability tendencies}. Each node represents a scientific publication, and an edge is established between two nodes if both articles were tweeted by at least one user. For the resulting network, we are considering the top $N=2,000$ nodes and $E=990,930$ edges. For clarity of the visualisation, only $8,000$ edges are displayed. Node size represents the total number of users who mentioned the corresponding article in a post. While node colour indicates if these users are more prone to share reliable or unreliable content, normalised by the total ($i.e.$, the node size). For visualisation purposes, the colour scale is truncated to the interval [$0.8, 1$], indicating that in general, users are more prone to share scientific content.}
    \label{fig:map_unreliability}
\end{figure}

\begin{figure}[H]
    \centering
    \includegraphics[width=1\linewidth]{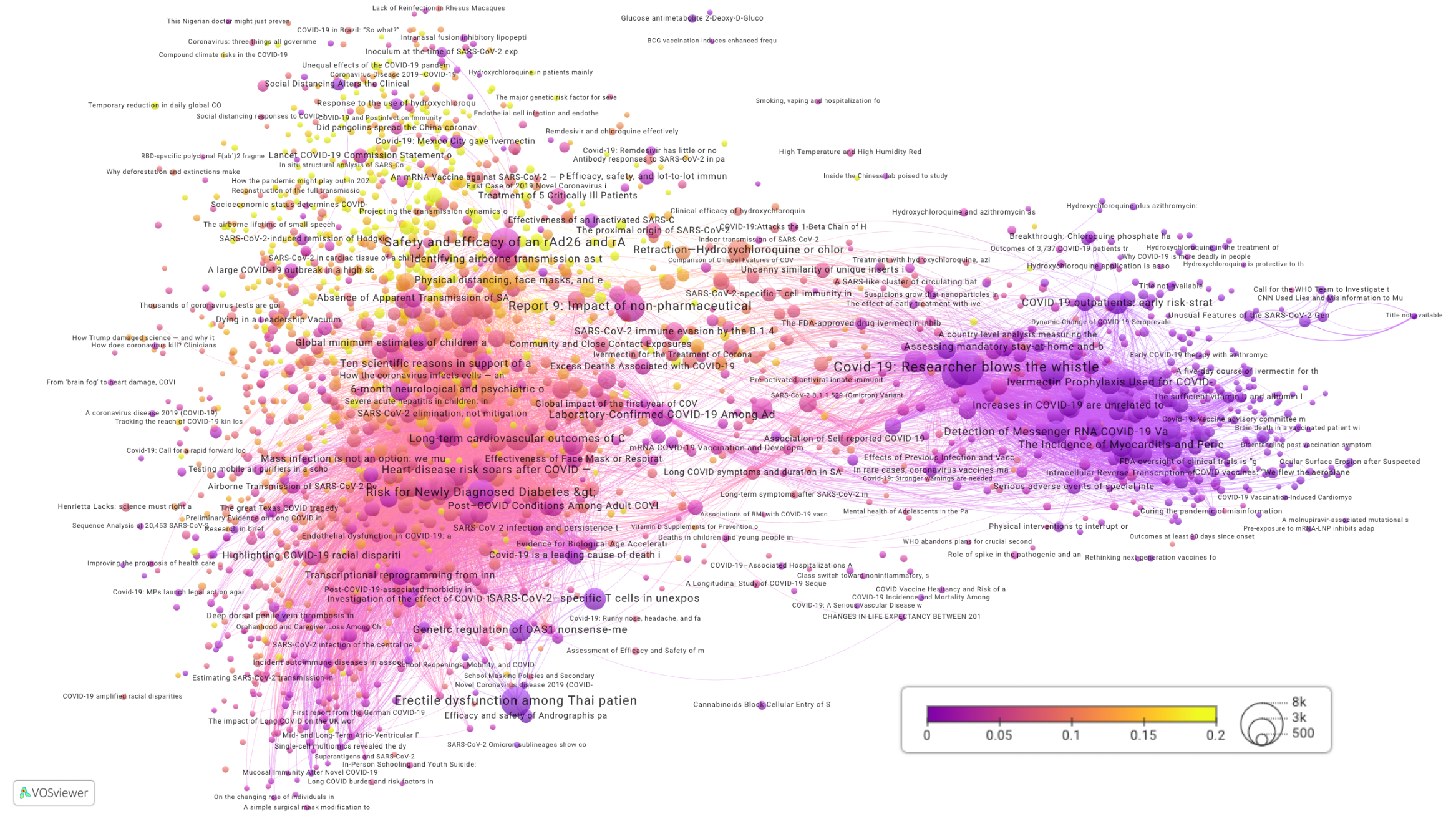}
    \caption{{\bf Network of publications co-shared by scientists and non-scientist users}. Each node represents a scientific publication, and an edge is established between two nodes if both articles were tweeted by at least one user. For the resulting network, we are considering the top $N=2,000$ nodes and $E=990,930$ edges. For clarity of the visualisation, only $8,000$ edges are displayed. Node size represents the total number of users who mentioned the corresponding article in a post. While node colour indicates the proportion of these users that are scientists, normalised by the total ($i.e.$, the node size). Thus, colours range from violet ($0$), indicating no scientists' presence, to yellow ($1$), indicating exclusive presence of scientists. For visualisation purposes, the colour scale is truncated to the interval [$0, 0.2$].}
    \label{fig:map_sci}
\end{figure}

%%%%%%%%%%%%%%%%%%%%%%%%%%%%%%%%%%%%%%%%%%%%%%%%%%%%%%%%%%%%%%%%%%%%
%%%%%%%%%%%%%%%%%%%%%%%%%%%%%%%%%%%%%%%%%%%%%%%%%%%%%%%%%%%%%%%%%%%%
%%%%%%%%%%%%%%%%%%%%%%%%%%%%%%%%%%%%%%%%%%%%%%%%%%%%%%%%%%%%%%%%%%%%
%%%%%%%%%%%%%%%%%%%%%%%%%%%%%%%%%%%%%%%%%%%%%%%%%%%%%%%%%%%%%%%%%%%%
%%%%%%%%%%%%%%%%%%%%%%%%%%%%%%%%%%%%%%%%%%%%%%%%%%%%%%%%%%%%%%%%%%%%
\subsection*{Scientists profiles}

\begin{figure}[H]
    \centering
    \includegraphics[width=1\linewidth]{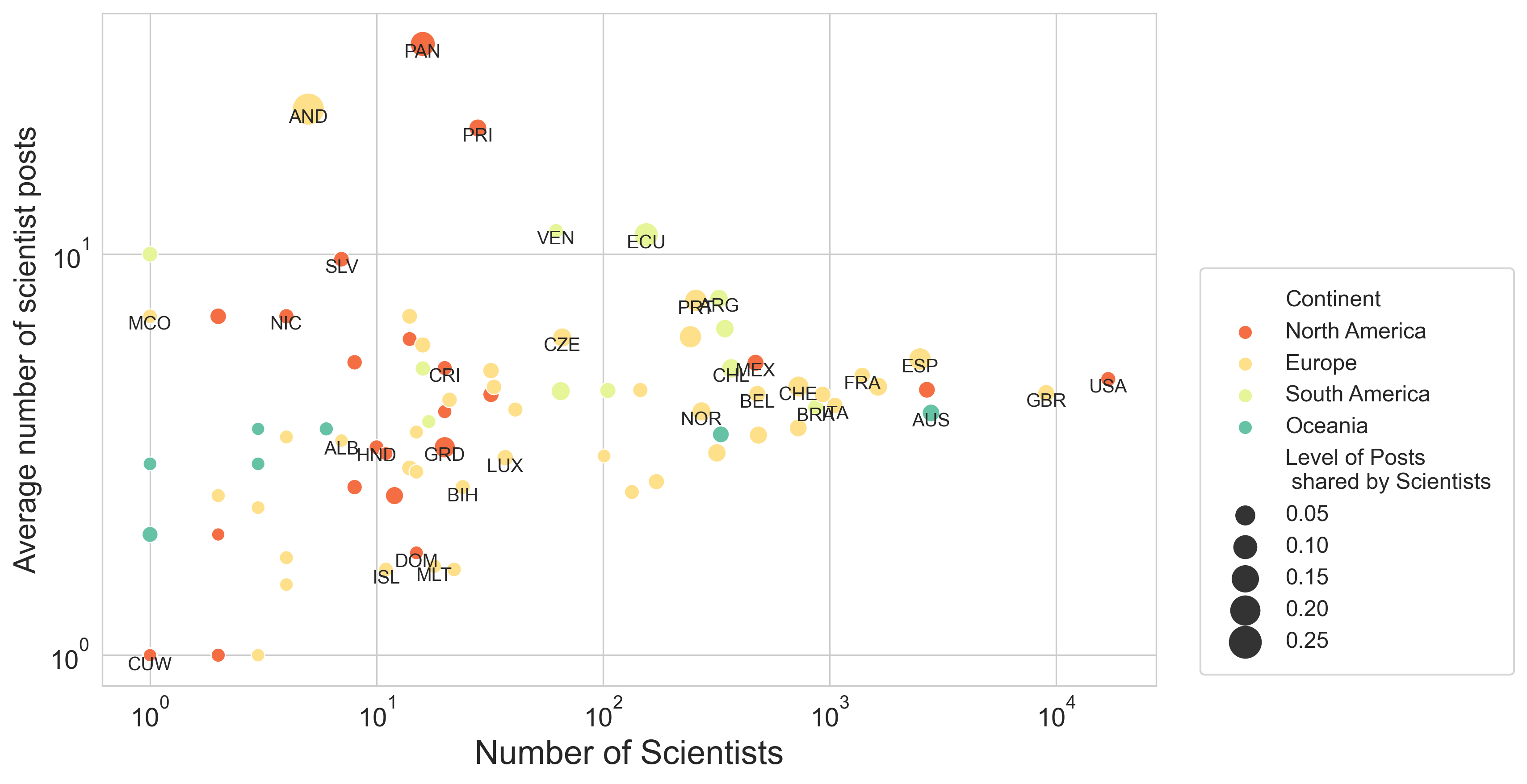}
    \caption{{\bf Scientists productivity as a function of the number of scientists present on Twitter per country}. The y-axis is measured as the number of posts generated by scientists, normalised by the number of scientists. The x-axis refers to the total of unique scholarly accounts. The size of each dot represents the level of posts shared by scientists, normalised by the total number of users (including non-scientist accounts). Notice that here we are only considering users who shared science and untrustworthy content at least once.}
    \label{fig:avg_prod_vs_num_scientists}
\end{figure} 

\end{document}